\documentclass[11pt,a4paper,dvipsnames,UTF8]{article}
\usepackage{CJK}
\usepackage{graphicx,color} 
\graphicspath{ {images} }
\usepackage{jheppub}
\usepackage{amsmath}
\usepackage{amssymb}
\usepackage{caption}     
\usepackage{mathtools}
\usepackage{comment}
\usepackage{tabularx}
\usepackage{booktabs}
\usepackage[utf8]{inputenc}
\usepackage{parskip}
\usepackage{physics}
\usepackage{tikz}
\usetikzlibrary{calc,arrows.meta,decorations,decorations.pathmorphing,positioning,decorations.markings}

\newcommand{\overbar}[1]{\mkern 1.5mu\overline{\mkern-1.5mu#1\mkern-1.5mu}\mkern 1.5mu}

\def\D{\Delta}

\def\f{\phi}
\def\db{\overbar{D}}

\def\zb{\bar{z}}

\def\x{\mathbf{x}}

\def \betal_1{\frac{ (r_--r_+)-2  (r_-\tanh ^{-1}\frac{r}{{r_-}}-r_{+}\tanh ^{-1}\frac{r}{{r_{+}}})}{2(r^2_--r^2_+)}}
\newcommand{\nn}{\nonumber\\}

\newmuskip\pFqmuskip

\newcommand*\pFq[6][8]{%
  \begingroup 
  \pFqmuskip=#1mu\relax
  \mathchardef\normalcomma=\mathcode`,
  \mathcode`\,=\string"8000
  \begingroup\lccode`\~=`\,
  \lowercase{\endgroup\let~}\pFqcomma
  {}_{#2}F_{#3}{\left[\genfrac..{0pt}{}{#4}{#5};#6\right]}%
  \endgroup
}
\newcommand{\pFqcomma}{{\normalcomma}\mskip\pFqmuskip}

\title{de Sitter locality from conformal field theory}
\author[a]{Parijat Dey,}
\author[b]{Zhongjie Huang (黄中杰)}
\author[c]{and Arthur Lipstein }

\affiliation[a]{ Department of Astrophysics and High Energy Physics,\\
S.N. Bose National Centre for Basic Sciences,\\
Salt Lake, Kolkata 700106, India}
\affiliation[b]{Zhejiang Institute for Modern Physics, Department of Physics,\\ Zhejiang University, Hangzhou,
Zhejiang 310058, China
}
\affiliation[c]{Department of Mathematical Sciences, Durham University,\\
Stockton Road, DH1 3LE, Durham, United Kingdom}
\emailAdd{parijat.dey@bose.res.in}
\emailAdd{zjhuang@zju.edu.cn}
\emailAdd{arthur.lipstein@durham.ac.uk }

\abstract{
An important insight from the study of AdS/CFT is that bulk locality can be derived from crossing symmetry of the boundary CFT. In this paper, we take the first steps in extending this statement to de Sitter background by demonstrating how to reconstruct a conformally coupled scalar effective field theory (EFT) with higher derivative interactions in four-dimensional de Sitter space from its in-in correlators. The latter can be computed from a certain EFT in Euclidean Anti-de Sitter space involving two scalar fields, which we derive from crossing symmetry of boundary correlators along with two novel constraints arising from unmixing anomalous dimensions of degenerate operators and equating them in different OPE channels. To facilitate the analysis, we work in Mellin space and apply dispersion relations to extract anomalous dimensions more efficiently.
}

\begin{document}

\begin{CJK*}{UTF8}{}
\CJKfamily{gbsn}
\maketitle
\end{CJK*}

\section{Introduction}

A very fruitful approach to the study of quantum gravity is the holographic principle, which relates it to a non-gravitational theory residing in the boundary of spacetime. This principle has been successfully applied in Anti-de Sitter background, where there is a dual formulation in terms of a conformal field theory (CFT) in the boundary \cite{Maldacena:1997re}, but its generalisation to more realistic backgrounds such as those relevant to cosmology is still a major open question \cite{Maldacena:2002vr,McFadden:2009fg,Maldacena:2011nz,Anninos:2011ui}. According to the inflationary paradigm, the early Universe had an approximately de Sitter geometry \cite{Guth:1980zm,Linde:1981mu,Albrecht:1982wi}, so this will be the main focus of our attention.  

One of the earliest triumphs of the AdS/CFT correspondence was the derivation of bulk locality from crossing symmetry of the boundary CFT correlators \cite{Heemskerk:2009pn}. Later work built on this in various ways using powerful tools of the conformal bootstrap \cite{Ferrara:1973yt,Ferrara:1972kab,Polyakov:1974gs}, although much of this progress relied heavily on supersymmetry (see for example \cite{Bissi:2022mrs} for a recent review). By comparison, holography in de Sitter background is far less established, although in recent years there has been exciting progress in the cosmological  bootstrap \cite{Baumann:2022jpr}, to a large extent driven by the study of factorisation properties \cite{Arkani-Hamed:2018kmz,Goodhew:2020hob}. While general solutions to the conformal Ward identities for cosmological correlators have been found for an arbitrary number of points \cite{Bzowski:2013sza,Bzowski:2020kfw}, the constraints of conformal symmetry have yet to be fully leveraged with those of factorisation.

In this paper, we will investigate whether bulk locality in four-dimensional de Sitter space (dS$_4$) can be derived from boundary conformal field theory. In de Sitter space, the observables are in-in correlators, which can be computed via the Schiwnger-Keldysh formalism \cite{Weinberg:2005vy} or via expectation values obtained by squaring the cosmological wavefunction \cite{Maldacena:2002vr}. On the other hand, it was recently shown that in-in correlators can be computed in terms of Witten diagrams in Euclidean AdS (EAdS) \cite{Sleight:2020obc,Sleight:2021plv}, which in the case of scalar correlators can be derived from an effective field theory (EFT) with double the number of scalar fields in EAdS \cite{DiPietro:2021sjt}. The scalar fields come in pairs, which are formally dual to boundary operators related by a shadow transformation. We shall refer to this as the shadow formalism.      

Our strategy will be to start with a general scalar effective field theory in dS$_4$ consisting of local four-point interactions with an arbitrary number of derivatives (modulo equations of motion and integration by parts). Using the shadow formalism, we will then map this to a scalar EFT in EAdS$_4$ with double the number of scalar fields. Note that the resulting EFT contains non-trivial interactions between the two scalar fields. We will then demonstrate how to reconstruct this EFT from the boundary conformal correlators. In contrast to the usual story found long ago in AdS, we find that crossing symmetry is not enough to reconstruct the shadow EFT. In particular, we must supplement it with two additional constraints. First we must unmix the anomalous dimensions of operators which are classically degenerate and demand that half of them vanish (we shall refer to this as the unmixing constraint). Then we compute the non-vanishing anomalous dimensions in different channels of the operator product expansion and equate them (we will refer to this as the cross-channel constraint). Interestingly, the procedure we use for unmixing degenerate operators was first developed in the context of AdS$_5$/CFT$_4$ \cite{Aprile:2017xsp}. 

For simplicity, we only consider conformally coupled scalars in the bulk since their correlators do not exhibit infrared divergences in contrast to the physically more relevant case of massless scalars, which we discuss further in the conclusion and leave for future work. Another technical point worth mentioning is our use of Mellin space to do CFT calculations \cite{Mack:2009mi,Mack:2009gy,Penedones:2010ue,Fitzpatrick:2011ia}, which has major advantages over position space in our context since three-dimensional conformal blocks do not have a closed form in position space \cite{Caron-Huot:2020ouj}. In particular, we develop dispersion relations in Mellin space which allow us to extract anomalous dimensions much more efficiently than in position space. Mellin space methods for cosmological correlators were previously developed in \cite{Sleight:2019mgd,Sleight:2020obc}. For other recent applications of conformal bootstrap ideas to cosmology, see for example \cite{Hogervorst:2021uvp,Loparco:2023rug,Bissi:2023bhv}.

This paper is organised as follows. In Section \ref{oview}, we review some basic concepts of the conformal bootstrap and how bulk locality was derived from crossing symmetry in the context of AdS/CFT. We then review the definition of in-in correlators in de Sitter space and how to compute them using the shadow formalism, and sketch the main results of the paper. In Section \ref{mellincft}, we review the conformal bootstrap in Mellin space and describe dispersion relations for extracting anomalous dimensions. In Section \ref{boundary} we spell out our bootstrap procedure for reconstructing the bulk shadow EFT from boundary CFT correlators. We then match the results obtained from this procedure with explicit bulk calculations in Section \ref{bulk}. We present our conclusions and future directions in Section \ref{conclusion}. There are also a number of Appendices with further technical details. In Appendix \ref{appx:mack}, we provide more details on Mack polynomials which are useful for performing Mellin space calculations. In Appendix \ref{unmixing} we derive a useful formula for unmixing anomalous dimensions. In Appendix \ref{spin2} we provide more details on results of bootstrapping in Mellin space. In Appendix \ref{positionsp} we provide details of bulk calculations, and in Appendix \ref{dmellin} we explain how to convert them to Mellin space.

\section{Overview} \label{oview}

In this section, we will first review some basic concepts of conformal field theory and how crossing symmetry can be used to derive a local scalar effective field theory in EAdS. We then review cosmological correlators and how they can be computed using an effective action in EAdS, and sketch how to reconstruct this action from boundary conformal field theory, which will be the main goal of this paper.

\subsection{AdS locality from CFT} \label{adscf_review}
We will now review some aspects of CFT in position space and how it can be used to derive bulk locality. A CFT is specified by a set of local operators $\mathcal{O}_{\D,\ell}$, labeled by their scaling dimensions $\D$ and spins $\ell$. The quantity $\tau=\D-\ell$ is called the twist of the operator $\mathcal{O}$. The product of two operators in a CFT can be written as an infinite sum of local operators with unfixed coefficients, known as OPE coefficients. The scaling dimensions and OPE coefficients together form the CFT data. The power of conformal symmetry lies in the fact that it fixes the two- and three-point correlators. The dynamical information of a CFT is therefore encoded in four-point correlators. Given four  points $x_i^\mu\, , i=1,\cdots 4\,,$ one can construct conformally invariant cross-ratios $z, \zb$ or $U, V $ defined as \footnote{We will use $z, \zb$ and $U, V$ interchangeably in what follows.}
\begin{align}
U= z \zb=\frac{x^2_{12}x^2_{34}}{x^2_{13}x^2_{24}}\,,\quad V= (1-z)(1- \zb)=\frac{x^2_{23}x^2_{14}}{x^2_{13}x^2_{24}}\,.
\end{align}
The correlation function of four scalar  operators can be written as a function of cross-ratios 
\begin{align}\label{4ptfn}
\langle {\f_1(x_1)\f_2}(x_2) {\f_3(x_3)\f_4}(x_4)\rangle=\frac{1}{(x^2_{12})^{\frac{\D_1+\D_2}{2}}(x^2_{34})^{\frac{\D_3+\D_4}{2}}}\left(\frac{x^2_{14}}{x^2_{24}}\right)^a \left(\frac{x^2_{14}}{x^2_{13}}\right)^b \mathcal{G}_{\{\D_i\}}(z, \zb)
\end{align}
where $a=\frac{\D_2-\D_1}{2}, b=\frac{\D_3-\D_4}{2}$ and $\{\D_i\}$ is the shorthand for $\{\D_1,\D_2,\D_3,\D_4\}$. The function $\mathcal{G}_{\{\D_i\}}(z, \zb)$ admits the following expansion in terms of the  CFT data \cite{Dolan:2000ut}:
\begin{align}\label{partialwaves}
\mathcal{G}_{\{\D_i\}}(z, \zb) &= \sum_{\D, \ell}{a}_{\D, \ell}\ g^{a,b}_{\D, \ell }(z,\zb)
\end{align}
where 
\begin{align}
{a}_{\D,\ell}=C_{12{\mathcal{O}}} C_{34 \mathcal{O}}
\end{align}
are the square of OPE coefficients appearing in the s-channel \footnote{We associate the $s,t,u$ channels to the partitions $(1,2;3,4),(1,4;2,3)$ and $(1,3;2,4)$, respectively.} and $g^{a,b}_{\D,\ell}$ are conformal blocks, which satisfy a differential equation obtained from the conformal Casimir \cite{Dolan:2003hv, Dolan:2011dv}. The correlator \eqref{4ptfn} also admits an expansion in the $t$-channel which can be obtained from  \eqref{4ptfn} by the interchanging $2\leftrightarrow 4$. The associativity of the OPE implies that the $s$-channel and $t$-channel expansions of the correlator should be the same and results in the following crossing equation:
\begin{align}\label{booteq1}
\mathcal{G}_{\{\D_i\}}(z, \zb)
&=\frac{(x^2_{12} x^2_{34})^{\frac{\D_3+\D_4}{2}}}{(x^2_{14}x^2_{23})^{\frac{\D_2+\D_3}{2}}}(x^2_{13}x^2_{24})^{\frac{\D_2-\D_4}{2}}\mathcal{G}_{\{\D_i\}}(1-z, 1-\zb).
\end{align}
The goal of the conformal bootstrap program is to solve \eqref{booteq1} and compute the CFT data $\{\D, C_{\D, \ell}\}$. See \cite{Rattazzi:2008pe,Fitzpatrick:2012yx, Komargodski:2012ek, Gopakumar:2016cpb, Alday:2016njk, Caron-Huot:2017vep, Bissi:2019kkx, Bissi:2021spj} and references therein for details.

In this work we are interested in computing correlators of a three dimensional CFT living in the boundary of dS$_4$. A closed form expression for 3d the conformal blocks is not known but they can be written as a sum over 2d blocks  and admit a series expansion  around $z=0$ given below \cite{Hogervorst:2016hal, Caron-Huot:2020ouj}:
\begin{align}
g^{a,b}_{\D, \ell }(z,\zb)&=\left(1-\frac{z}{\zb}\right)^{-\frac{1}{2}}\sum_{m=0}^{\infty}z^{\frac{\D-\ell}{2}+m} h^{(m)}_{\D,\ell}(\zb)\,,\nn
\text{with} \quad h^{(m)}_{\D,\ell}(\zb) &=\sum_{i=-m}^m  h^{(m,i)}_{\D,\ell} k^{(a,b)}_{\D+\ell+2i}(\zb)\,,\nn
 \text{and}\quad k^{(a,b)}_{\beta}(z) &=z^{\frac{\beta}{2}}\, _2F_1(\beta/2+a,\beta/2+b,\beta,z)\,,
\end{align}
where $_2F_1$ is the Gaussian hypergeometric function. There exists a recursion relation in $m$ \cite{Caron-Huot:2020ouj} that allows us to compute the conformal blocks for each value of $m$. This results in an expansion of the conformal blocks in powers of $z$ to all orders. 

Let us now review how the authors of \cite{Heemskerk:2009pn} derived AdS locality from crossing symmetry. First they considered 4-point correlators of identical scalar operators in an even-dimensional CFT which admits an expansion in a central charge $c$. They then expanded the CFT data to first non-trivial order in the central charge expansion as follows: 
\begin{align}
C_{n,\ell}=&\ C_{n,\ell}^{(0)}+\frac{1}{c}C_{n,\ell}^{(1)}+\cdots,\\
\Delta=&\ 2\Delta_{0}+2n+\ell+\frac{1}{c}\gamma^{(1)}_{n,\ell}+\cdots,
\end{align}
where $\Delta_0$ is referred to as the classical scaling dimension and $\gamma$ is the anomalous dimension. They then plugged this into the crossing equation \eqref{booteq1} and truncated the sum over spin in the conformal block expansion in \eqref{partialwaves}. After doing so they were able to derive a recursion relation for the the anomalous dimensions and found that the number of solutions with spin at most $L$ is $(L+2)(L+4)/8$, which is in one-to-one correspondence with local 4-point interactions of a massive scalar field in EAdS, modulo equations of motion and integration by parts. Note that odd-spin interactions can be removed using equations of motion and integration by parts. For example, 2-derivative interactions (which would correspond to spin-1) can be reduced to a zero-derivative interaction in this way. In more detail, there are $L/2+1$ independent interactions corresponding to spin-$L$, with the number of derivatives ranging from $2L$ to $3L$ (in intervals of two). For example, there is one interaction vertex corresponding to spin-0 and two vertices corresponding to spin-2, with 4 derivatives and 6 derivatives, respectively. Hence the total number of bulk interaction vertices with spin at most $L$ is $\sum_{\ell=0}^{L/2}\left(l+1\right)=(L+2)(L+4)/8$. 

In summary, \cite{Heemskerk:2009pn} used crossing symmetry of boundary CFT correlators to reconstruct a bulk effective action in EAdS$_{d+1}$ of the form 
\begin{equation}
S=\int \frac{dZ d^{d}x}{Z^{d+1}}\left[\frac{1}{2}\left(\partial\phi\cdot\partial\phi-m^{2}\phi^{2}\right)-V(\phi,\cdots)\right],
\label{adsaction}
\end{equation}
where the potential describes 4-point interactions and we use the Poincare patch metric with unit radius
\begin{equation}
ds^{2}=\frac{d\vec{x}^{2}+dZ^{2}}{Z^{2}},
\label{adsmetric}
\end{equation}
where $0\leq Z \leq \infty$ and $\vec{x}$ is a vector in the $d$-dimensional boundary. Note that derivatives in \eqref{adsaction} are contracted with the flat metric, and the potential depends on the scalar field acted on by an arbitrary number of derivatives, which is denoted by the ellipsis. In more detail the potential has the form
\begin{equation}
V(\phi)=\lambda_{0}\phi^{4}+\lambda_{4}\left(\partial\phi\right)^{4}+\lambda_{6}\left(\partial\phi\right)^{2}\left(\nabla^{\mu}\nabla_{\mu}\phi\right)^{2}+\cdots,
\label{adspot}
\end{equation}
where the coefficients are unfixed and the ellipsis denote higher-derivative terms. We have only listed terms corresponding to spin-0 and spin-2 interactions (recall that spin-2 interactions can have four or six derivatives, as mentioned above). 

The counting of bulk interaction vertices can be understood by analogy with flat space S-matrices. In particular, 4-point S-matrices can be constructed from the following basis:
\begin{equation}
(st)^{L/2}u^{c},\,\,\,c\in\left\{ 0,\cdots,L/2\right\} 
\end{equation}
where $s,t,u$ are the usual Mandelstam variables satisfying $s+t+u=4 m^{2}$. These basis elements correspond to interaction vertices of the form
\begin{equation}
\nabla_{\mu_{1}\cdots\mu_{L/2}\nu_{1}\cdots\nu_{L/2}\rho_{1}\cdots\rho_{c}}\phi\nabla_{\mu_{1}\cdots\mu_{L/2}}\phi\nabla_{\rho_{1}\cdots\rho_{c}}\phi\nabla_{\nu_{1}\cdots\nu_{L/2}}\phi.
\end{equation}
The analogy to flat space amplitudes can be made precise using Mellin space, and greatly simplifies the counting of solutions to the crossing equations from the boundary point of view, as we will explain in Section \ref{mellinloc}. The use of Mellin space has additional technical advantages for CFT calculations in odd dimensions, since conformal blocks in position space are extremely complicated in odd dimensions.

\subsection{de Sitter correlators} \label{dscorrelators}

Now let us consider de Sitter background. We will work in the Poincare patch
\begin{equation}
ds^{2}=\frac{d\vec{x}^{2}-d\eta^{2}}{\eta^{2}},
\end{equation}
where $-\infty<\eta<0$ is the conformal time, $\vec{x}$ is a $d$-dimensional vector in the boundary, and we set radius to 1. The future boundary is located at $\eta=0$. The observables in this background are known as in-in correlators. For scalar fields, these are defined as follows:
\begin{equation}
\left\langle \phi\left(\vec{x}_{1},\eta\right)\cdots\phi\left(\vec{x}_{n},\eta\right)\right\rangle =\frac{\left\langle 0\right|U^{\dagger}(-\infty,\eta)\phi\left(\vec{x}_{1},\eta\right)\cdots\phi\left(\vec{x}_{n},\eta\right)U(-\infty,\eta)\left|0\right\rangle }{\left\langle 0\right|U^{\dagger}(-\infty,\eta)U(-\infty,\eta)\left|0\right\rangle },
\end{equation}
where $\left|0\right\rangle $ us the Bunch-Davies vacuum and $U$ is the evolution operator in the interaction picture. In practice we will consider correlators of operators on the future boundary so take $\eta\rightarrow0$. In-in correlators can be computed using the Schwinger-Keldysh formalism which involves time-ordered and anti-time-ordered propagators \cite{Maldacena:2002vr,Weinberg:2005vy}, or using Witten diagrams in EAdS using the shadow formalism \cite{Sleight:2020obc,Sleight:2021plv,DiPietro:2021sjt}, which we will briefly review below. For a more a detailed discussion see \cite{Heckelbacher:2022hbq}.

We will consider a general EFT of the form \eqref{adsaction} and \eqref{adspot} in de Sitter space. We then define time-ordered and anti-time-ordered fields (denoted by the subscript $T$ and $A$, respectively) with the following action: 
\begin{equation}
S_{\mathrm{SK}}=\int\frac{d^{d}xd\eta_{T}}{\eta_{T}^{d+1}}\mathcal{L}\left(\phi_{T},\cdots\right)-\int\frac{d^{d}xd\eta_{A}}{\eta_{A}^{d+1}}\mathcal{L}\left(\phi_{A},\cdots\right),
\end{equation}
where the Lagrangians  for the two scalar fields are identical to the one in \eqref{adsaction} and we include an ellipsis since the potential can depend on scalar fields as well any number of covariant  derivatives acting on them. We then Wick rotate to EAdS using the following prescription:
\begin{equation}
\eta_{T}\rightarrow-i\,Z,\,\,\,\eta_{A}\rightarrow i\,Z,
\end{equation}
where we recall that $0<Z<\infty$, and perform the following field redefinitions:
\begin{equation}
\phi_{A}=e^{-i\frac{\pi}{2}\Delta_{+}}\phi_{+}+e^{-i\frac{\pi}{2}\Delta_{-}}\phi_{-},\,\,\,\phi_{T}=e^{i\frac{\pi}{2}\Delta_{+}}\phi_{+}+e^{i\frac{\pi}{2}\Delta_{-}}\phi_{-},
\end{equation}
where the scaling dimensions are related to the mass of the scalars in the standard way:
\begin{equation}
\Delta_{\pm}=\frac{d}{2}\pm\sqrt{\frac{d^{2}}{4}+m^{2}}.
\label{deltamass}
\end{equation}
For simplicity, we will restrict to conformally coupled scalars which correspond to
\begin{equation}
\Delta_{+}=\frac{d+1}{2},\,\,\,\Delta_{-}=\frac{d-1}{2}.
\end{equation}
Moreover we will set $d=3$, corresponding to a four-dimensional bulk. We then end up with the following action in EAdS$_4$, which we refer to as the shadow action:
\begin{equation}
iS_{\mathrm{shadow}}=\int\frac{dZ d^{3}x}{Z^{4}}\left[\frac{1}{2}\left(\left(\partial\phi_{+}\right)^{2}-m^{2}\phi_{+}^{2}\right)-\frac{1}{2}\left(\left(\partial\phi_{-}\right)^{2}-m^{2}\phi_{-}^{2}\right)-V\left(\phi_{+},\phi_{-}\right)\right],
\label{shadowaction}
\end{equation}
where covariant derivatives are understood to be contracted with the flat Euclidean metric and the first few terms in the potential are given by
\begin{equation}
V\left(\phi_{+},\phi_{-}\right)=V_{0}\left(\phi_{+},\phi_{-}\right)+V_{4}\left(\phi_{+},\phi_{-}\right)+V_{6}\left(\phi_{+},\phi_{-}\right)+\cdots,
\label{shadowpot1}
\end{equation}
with
\begin{align}
V_{0}\left(\phi_{+},\phi_{-}\right)&=\lambda_{0}\left(\phi_{+}^{4}-6\phi_{+}^{2}\phi_{-}^{2}+\phi_{-}^{2}\right),   \nonumber \\
V_{4}\left(\phi_{+},\phi_{-}\right)&=\lambda_{4}\left[\left(\partial\phi_{+}\right)^{4}+\left(\partial\phi_{-}\right)^{4}-2\left(\partial\phi_{+}\right)^{2}\left(\partial\phi_{-}\right)^{2} -4\left(\partial\phi_{+}\cdot\partial\phi_{-}\right)^{2} \right],  \nonumber    \\
V_{6}\left(\phi_{+},\phi_{-}\right)&=\lambda_{6}\left[\left(\partial\phi_{+}\right)^{2}\left(\nabla^{\mu}\nabla^{\nu}\phi_{+}\right)^{2}+\left(\partial\phi_{+}\right)^{2}\left(\nabla^{\mu}\nabla^{\nu}\phi_{+}\right)^{2}\right. \nonumber \\
& \,\,\,\,\,\,\,\,\,\,\,\,     -\left(\partial\phi_{+}\right)^{2}\left(\nabla^{\mu}\nabla^{\nu}\phi_{-}\right)^{2}-\left(\partial\phi_{-}\right)^{2}\left(\nabla^{\mu}\nabla^{\nu}\phi_{+}\right)^{2}  \nonumber \\
& \,\,\,\,\,\,\,\,\,\,\,\, \left.-4\left(\partial\phi_{+}\cdot\partial\phi_{-}\right)\left(\nabla^{\mu}\nabla^{\nu}\phi_{+}\right)\left(\nabla_{\mu}\nabla_{\nu}\phi_{-}\right)\right].
\label{shadowpot2}
\end{align}
Note the $\phi_{\pm}$ fields interact non-trivially with each other.

Now we ask the following question: can we reconstruct the effective action above from boundary CFT correlators? Since there are two types of bulk fields corresponding to two types of boundary operators, crossing symmetry of 4-point boundary correlators admits more solutions than in the case of a single bulk scalar field described in the previous subsection. For example, we find odd-spin solutions to the truncated crossing equations, which can arise from bulk interactions of the form
\begin{equation}
V\left(\phi_{+},\phi_{-}\right)=\partial\phi_{+}\cdot\partial\phi_{+}\phi_{-}^{2}.
\label{spin1}
\end{equation}
Note that such an interaction does not appear in \eqref{shadowpot2}. Using the analogy to flat space amplitudes discussed in the previous subsection, we see that this gives an amplitude proportional to $s$, corresponding to a spin-1 interaction. Similarly, we obtain solutions to the truncated crossing equations corresponding to higher-derivative bulk interactions with odd spin. This will be spelled out in Section \ref{mellincrossing}. In order to reproduce the effective action in \eqref{shadowaction}-\eqref{shadowpot2}, we must therefore impose additional constraints on the boundary correlators beyond crossing symmetry.   

\begin{figure}[h]
\centering
\begin{tikzpicture}[scale=1.5]
\draw[red,thick]  (1,0.5) -- (0,0.5) node [text width=2 cm,midway,above,align=center ] {$\D, \ell$} ;
\filldraw[black] (1,0.5) circle (1pt) ;
\filldraw[black] (0,0.5) circle (1pt) ;
\draw   (0,0.5) -- (-0.5,1) node[left] {$+$};
\filldraw[black] (-0.5,1) circle (1pt) ;
\draw  (0,0.5)-- (-0.5,0)node[left ] {$+$} ;
\filldraw[black] (-0.5,0) circle (1pt) ;
\draw   (1,0.5) -- (1.5,1) node[right]{$-$};
\filldraw[black] (1.5,1) circle (1pt) ;
\draw   (1,0.5) -- (1.5,0) node[right]{$-$};
\filldraw[black] (1.5,0) circle (1pt) ;
\draw (0,-1)  node[anchor=west] {s-channel};
\draw (4.5,-1)  node[anchor=west] {t-channel};
\filldraw[black] (5,1) circle (1pt) ;
\filldraw[black] (5,0.3) circle (1pt) ;
\filldraw[black] (4,0) circle (1pt) ;
\filldraw[black] (6,0) circle (1pt) ;
\draw  (5,0.3)-- (6,0) node[right]{$-$};
\draw   (5,0.3)-- (4,0)node[left ] {$+$} ;
\draw[red,thick]  (5,0.3)-- (5,1) node[anchor=north west] {$\Delta,\ell$};
\filldraw[black] (6,1.5) circle (1pt) ;
\draw   (5,1)-- (6,1.5) node[right]{$-$};;
\filldraw[black] (4,1.5) circle (1pt) ;
\draw    (5,1) -- (4,1.5) node[left ] {$+$};
\end{tikzpicture}
\caption{Equate anomalous dimensions in two different OPE channels}\label{fig:crossing} 
\end{figure}
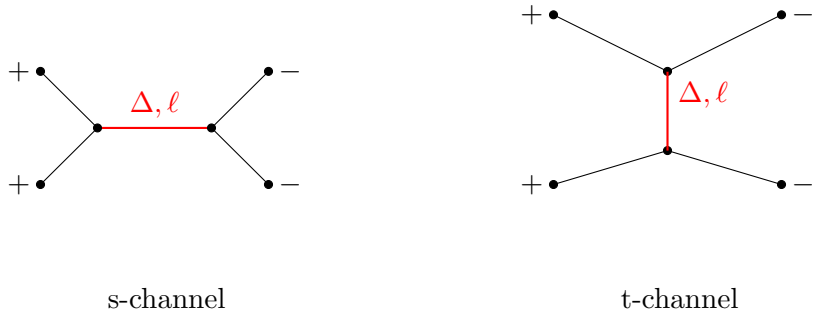

To understand the origin of these additional constraints, we consider three types of 4-point correlators: $\left\langle ++++\right\rangle $, $\left\langle ----\right\rangle $, and $\left\langle ++--\right\rangle $, where $\pm$ refers to the operators $\mathcal{O}_{\pm}$ which are dual to $\phi_{\pm}$, respectively. The following double-trace operators then appear in the conformal block expansion of these correlators:
\begin{align}
    :\mathcal{O}_- \square^n \partial^\ell \mathcal{O}_-:,\quad :\mathcal{O}_+ \square^n \partial^\ell \mathcal{O}_+:,\quad :\mathcal{O}_+ \square^n \partial^\ell \mathcal{O}_-:.
\end{align}
Note that $:\mathcal{O}_{+}\square^{n}\partial^{\ell}\mathcal{O}_{+}:$ and $:\mathcal{O}_{-}\square^{n+1}\partial^{\ell}\mathcal{O}_{-}:$ are classically degenerate and therefore mix together. In the quantum theory, they acquire anomalous dimensions and can be unmixed using a procedure first developed on the context of maximal super-Yang-Mills \cite{Aprile:2017xsp} and described in more detail in Section \ref{sec:adcon} and Appendix \ref{unmixing}. Let us denote the anomalous dimensions after unmixing as $\gamma^{\mathrm{mixed},\pm}_{n,l}$. We then impose that one of the anomalous dimensions is zero after unmixing:
\begin{equation}
\gamma^{\mathrm{mixed},-}_{n,l}=0.
\label{unmixingc}
\end{equation}
We refer to this as the unmixing constraint. Moreover, we denote the anomalous dimension of $:\mathcal{O}_{+}\square^{n}\partial^{\ell}\mathcal{O}_{-}:$ as $\gamma^{\mathrm{pure}}_{n,l}$ since it does not mix with other operators. We then impose the second constraint
\begin{equation}
\gamma^{\mathrm{mixed},+}_{n,l}=\gamma^{\mathrm{pure}}_{n+\frac{1}{2},l},
\label{crosschannel}
\end{equation}
which we refer to as the cross-channel constraint because it equates anomalous dimensions in two different OPE channels of the correlator $\left\langle ++--\right\rangle $, as depicted in Figure \ref{fig:crossing}. 
After imposing the constraints in \eqref{unmixingc} and \eqref{crosschannel}, which we collectively refer to as the anomalous dimension constraints, the resulting anomalous dimensions precisely match those obtained from boundary correlators computed from Witten diagrams using the effective action in \eqref{shadowaction}-\eqref{shadowpot2}. In particular, these additional constraints eliminate odd-spin interactions of the form \eqref{spin1} and fix the relative coefficients of the potential to match those in \eqref{shadowpot2}. Physically, these constraints encode that there is only one underlying scalar field in the original effective action in de Sitter space.

\section{CFT in Mellin space} \label{mellincft}

In this section, we review the Mellin representation of conformal correlators \cite{Mack:2009mi,Mack:2009gy,Penedones:2010ue,Fitzpatrick:2011ia}.  In Mellin space, conformal correlators are represented by Mellin amplitudes, which share many common features with scattering amplitudes in flat spacetime. This formalism is broadly used in the analytic bootstrap of holographic correlators in AdS, and will serve as a powerful tool for dS in-in correlators, since they can be written as EAdS Witten diagrams. For previous work on applying Mellin space to cosmological correlators, see \cite{Sleight:2019mgd,Sleight:2020obc}. After reviewing some properties of Mellin amplitudes, we will revisit the derivation of bulk locality in EAdS from crossing symmetry and derive dispersion relations for extracting anomalous dimensions from 4-point correlators more efficiently.

\subsection{Basic properties of Mellin amplitudes}

Compared to correlators in position space, Mellin amplitudes are usually simple functions with poles encoding exchange diagrams in the bulk and for bulk contact interactions, Mellin amplitudes are just polynomials like amplitudes in flat space. Moreover conformal blocks in Mellin space are analytic in the boundary spacetime dimension $d$. This property makes the Mellin space particularly suitable for the analytic study of conformal correlators in three dimensions.

The Mellin amplitude $\mathcal{M}$ of a four-point correlator \eqref{4ptfn}  is defined by the integral \footnote{The Mellin transform becomes apparent when the correlator is expressed in terms of the variables $U$ and $V$. For this reason, we work with $\mathcal{G}_{\{\Delta_i\}}(U,V)$ instead of $\mathcal{G}_{\{\Delta_i\}}(z,\zb)$.}
\begin{align}\label{eq:mellindef}
    \mathcal{G}_{\{\Delta_i\}}(U,V) =&\ \int_{-i\infty}^{i\infty} \frac{\dd s\, \dd t}{(2\pi i)^2} U^{\frac{s}{2}}V^{\frac{t-\Delta_2-\Delta_3}{2}}\mathcal{M}_{\{\Delta_i\}}(s,t)\, \Gamma_{\{\Delta_i\}}(s,t),
\end{align}
where $\Gamma_{\{\Delta_i\}}(s,t)$ is a product of $\Gamma$ functions 
\begin{align}
    \Gamma_{\{\Delta_i\}}(s,t) \equiv&\ \Gamma\left(\frac{\Delta_1+\Delta_2-s}{2}\right)\Gamma\left(\frac{\Delta_3+\Delta_4-s}{2}\right)\Gamma\left(\frac{\Delta_1+\Delta_4-t}{2}\right) \nn 
    &\times \Gamma\left(\frac{\Delta_2+\Delta_3-t}{2}\right)\Gamma\left(\frac{\Delta_1+\Delta_3-u}{2}\right)\Gamma\left(\frac{\Delta_2+\Delta_4-u}{2}\right),
\end{align}
and $s,t,u$ are Mellin-Mandelstam variables satisfying the constraint
\begin{align}
    s+t+u=\Delta_1+\Delta_2+\Delta_3+\Delta_4.
\end{align}
Only two of these variables are independent, which we will choose to be $s$ and $t$. In general, both the $\Gamma$ factors and the Mellin amplitude $\mathcal{M}$ have poles. These poles organize into left- or right-moving series. For instance, in $\Gamma\left(\frac{\Delta_1+\Delta_2-s}{2}\right)$ the poles are right-moving:
\begin{align}\label{eq:gammapole}
    \Gamma\left(\frac{\Delta_1+\Delta_2-s}{2}\right)\quad   \implies \quad \text{poles at } s=\Delta_1 + \Delta_2 +2n,\ n=0,1,2,...
\end{align}
The integration contour in \eqref{eq:mellindef} is defined such that it passes to the right (left) of all left-moving (right-moving) poles. To evaluate the integral \eqref{eq:mellindef}, one may deform the contour to encircle the poles moving in one direction. The integral then reduces to a sum over the residues at these poles, yielding a Taylor series expansion in the cross-ratios $U$ and $V$. This expansion can then be directly compared with the correlator $\mathcal{G}_{\{\Delta_i\}}(U,V)$. In general, we have the following correspondence between Mellin amplitudes and correlators in position space: 
\begin{align}
    \text{simple pole at } s=s_0 \quad \Longleftrightarrow \quad U^{\frac{s_0}{2}} \text{ in Taylor expansion}.
\end{align}

Crossing symmetry of Mellin amplitudes takes the same form as for scattering amplitudes. For example, exchanging operators 1 and 3 relates Mellin amplitudes as follows:
\begin{align}
    \mathcal{M}_{\Delta_1\Delta_2\Delta_3\Delta_4}(s,t) = \mathcal{M}_{\Delta_3\Delta_2\Delta_1\Delta_4}(t,s),
\end{align}
which is equivalent to the crossing equation \eqref{booteq1} in position space. Furthermore, if we consider a correlator with identical operators $\mathcal{O}$ of dimension $\Delta_{\phi}$, it becomes
\begin{align}
\mathcal{M}_{\Delta_{\phi}\Delta_{\phi}\Delta_{\phi}\Delta_{\phi}}(s,t) = \mathcal{M}_{\Delta_{\phi}\Delta_{\phi}\Delta_{\phi}\Delta_{\phi}}(t,s),
\end{align}

\subsection{Mack polynomials} 

The conformal block decomposition of position space correlators was reviewed in Section \ref{adscf_review}. In Mellin space, a conformal block with twist $\tau$  corresponds to a series of poles at $s = \tau + 2m$ with $m = 0,1,2,\dots$, which can be justified by expanding the conformal blocks in the variable $U$. The poles at $m > 0$ can be interpreted as arising from descendant operators. The residues at these poles are given by Mack polynomials $\mathcal{Q}$ \footnote{Note that both conformal blocks and Mack polynomials depend on the external dimensions $\{\Delta_i\}$ and the spacetime dimension $d$. We suppress these labels here and below to make the notation simpler. }  
\begin{align}\label{eq:block2Mack}
    g_{\tau,\ell}(U,V) \quad \Longleftrightarrow \quad \sum_{m=0}^{\infty} \frac{\mathcal{Q}_{\tau,\ell,m}(t)}{s - \tau - 2m},
\end{align}
and the conformal block decomposition \eqref{partialwaves} in Mellin space reads 
\begin{align}\label{eq:mexp}
    \mathcal{M}(s,t) \sim \sum_{\tau,\ell} \sum_{m=0}^{\infty} \frac{a_{\tau,\ell} \mathcal{Q}_{\tau,\ell,m}(t)}{s-\tau-2m}.
\end{align}
A closed-form formula for Mack polynomials with arbitrary $\tau$, $\ell$, $m$, external dimensions $\Delta_i$, and spacetime dimension $d$, was found in \cite{Costa:2012cb} and is given in Appendix \ref{appx:mack}. Let us point out two properties of Mack polynomials that will be important for our study: 
\begin{itemize}
    \item \textbf{Polynomial structure:} Mack polynomials are degree-$\ell$ polynomials in $t$. For example, the spin-1 Mack polynomial with identical external dimensions $\Delta_0$ takes the form
    \begin{align}
       \mathcal{Q}_{\tau,1,1}(t) = -\frac{2^{\tau} \tau\, \Gamma\left(\frac{\tau + 3}{2}\right) \left(2t - 4\Delta_0 + \tau + 2\right)}{\sqrt{\pi} (2\tau + 4 - d) \Gamma^3\left(\frac{\tau}{2} + 1\right) \Gamma^2\left(\frac{2\Delta_0 - (\tau + 2)}{2}\right)}.
    \end{align}
    Schematically, the exchange of operators is reflected in the analytic structure of Mellin amplitudes in the following way:
    \begin{align}\label{eq:block2Mack2}
        \text{$\mathcal{O}_{\tau,\ell}$ exchange in $s$-channel} \quad \Longleftrightarrow \quad \parbox{15em}{simple poles at $s = \tau, \tau+2, \dots$,\\degree-$\ell$ polynomial in $t$.}
    \end{align}

    \item \textbf{Vanishing for double-trace twists:} Mack polynomials vanish when $\tau$ is exactly double-trace, i.e., when $\tau = \Delta_1 + \Delta_2 + 2n$ or $\tau = \Delta_3 + \Delta_4 + 2n$. This stems from the factor
    \begin{align}
        \mathcal{Q}_{\tau,\ell,m}(t) \propto \frac{1}{\Gamma\left(\frac{\Delta_{1} + \Delta_{2} - (\tau + 2m)}{2}\right) \Gamma\left(\frac{\Delta_{3} + \Delta_{4} - (\tau + 2m)}{2}\right)},
    \end{align}
    in the expression \eqref{eq:mack1} of Mack polynomials, which introduces simple zeros at $\tau=\Delta_1 + \Delta_2 + 2n$ and $\tau=\Delta_3 + \Delta_4 + 2n$. When $\Delta_1 + \Delta_2 - \Delta_3 - \Delta_4 \in 2\mathbb{Z}$, these simple zeros coincide and enhance to double zeros. 
    
The underlying reason for this vanishing is that the poles associated with double-trace operators are already accounted for by the $\Gamma$ functions in $\Gamma_{\{\Delta_i\}}(s,t)$ (as we can see from \eqref{eq:gammapole}). As a result, conformal blocks with $\tau = \Delta_1 + \Delta_2 + 2n$ and $\tau = \Delta_3 + \Delta_4 + 2n$ do not introduce new poles in the Mellin amplitude.  Instead they contribute to regular terms with degree-$\ell$ polynomial behavior in $t$, while the poles are encoded by $\Gamma$ functions. In an interacting theory, double-trace operators will acquire anomalous dimensions and move away from $\tau=\Delta_1+\Delta_2+2n$ or $\tau=\Delta_3+\Delta_4+2n$. At first this seems to be a contradiction since the poles in $\Gamma_{\{\Delta_i\}}(s,t)$ remain fixed at the exact double-trace values. This is resolved by requiring Mellin amplitudes to vanish at these points
    \begin{align}
        \mathcal{M}(s,t)\big|_{s = \Delta_1 + \Delta_2 + 2n} = 0, \qquad \mathcal{M}(s,t)\big|_{s = \Delta_3 + \Delta_4 + 2n} = 0,
    \end{align}
    which is known as the Polyakov condition \cite{Penedones:2019tng,Caron-Huot:2020adz,Carmi:2020ekr}. When $\Delta_1 + \Delta_2 - \Delta_3 - \Delta_4 \in 2\mathbb{Z}$, we have double poles in the $\Gamma$ functions, and the Polyakov condition becomes  
    \begin{align}
        \mathcal{M}(s,t)\big|_{s = \Delta_1 + \Delta_2 + 2n} = 0, \qquad \partial_s \mathcal{M}(s,t)\big|_{s = \Delta_1 + \Delta_2 + 2n} = 0.
    \end{align}
\end{itemize}

\subsection{AdS locality from CFT in Mellin space} \label{mellinloc}

Let us now revisit the derivation of AdS locality from CFT using Mellin space techniques. The analysis in position space was first carried out in \cite{Heemskerk:2009pn} and reviewed in Section \ref{adscf_review}, and we will now show that it becomes dramatically simplified in Mellin space. In \cite{Heemskerk:2009pn}, the authors considered a single-scalar EFT in EAdS, whose boundary correlators involve identical scalar operators. At four points, the Mellin amplitude for such a correlator satisfies
\begin{align}\label{eq:crossingid}
    \mathcal{M}(s,t) = \mathcal{M}(t,s) = \mathcal{M}(s,u=4\Delta-s-t).
\end{align}
Since all external operators are identical, we omit the label $\{\Delta_i\}$ in $\mathcal{M}$. In Mellin space, the conformal block expansion of the correlator in the s-channel reads 
\begin{align}\label{eq:mexp1}
    \mathcal{M}(s,t) = \sum_{\tau,\ell} \sum_{m=0}^{\infty} \frac{a_{\tau,\ell} \mathcal{Q}_{\tau,\ell,m}(t)}{s-\tau-2m} + \text{(regular terms)},
\end{align}
where ``regular terms" stands for terms without $s$ poles, which encode operators with exactly double-trace twists (such operators appear when we are considering perturbative Mellin amplitudes). Following \cite{Heemskerk:2009pn}, if we truncate the sum over spin to $\ell \leq L$, due to the polynomial behavior of Mack polynomials, each term in \eqref{eq:mexp1} is a polynomial in $t$ with degree $\leq L$, and we can conclude that the whole function $\mathcal{M}(s,t)$ is also such a $t$ polynomial. Similarly, by considering the t-channel conformal block expansion, we conclude that $\mathcal{M}(s,t)$ is also a polynomial in $s$ with degree $\leq L$. Since $\mathcal{M}$ only depends on $s$ and $t$, it must be a polynomial as a whole. Furthermore, the crossing equation \eqref{eq:crossingid} implies that $\mathcal{M}$ should be a symmetric function in $s$, $t$ and $u$. Putting these conditions together, we have \footnote{These solutions might seem to violate the Polyakov condition. This is not a problem, since the Polyakov condition only holds for non-perturbative Mellin amplitudes.}
\begin{align}
    \mathcal{M}(s,t) = \text{(symmetric polynomial in $s,t,u$ with degree $\leq L$ in $s,t$)},
\end{align}
where we replace and $u$ with $4\Delta-s-t$ when counting the degree of the polynomials \footnote{In our counting, the degree of $s^m t^n$ is $\max(m,n)$, not $m+n$. }. There are no poles in the solution, which means that all operators in the OPE are double-trace operators, as we expect from the bulk picture. In fact, polynomials in Mellin space can be mapped to $D$-functions in position space, which are scalar contact diagrams in the bulk \cite{DHoker:1999kzh, Arutyunov:2002fh, Gary:2009ae} and the number of derivatives in the interactions vertices is directly related to the total degree of $\mathcal{M}$. 

We seem to magically obtain the result in \cite{Heemskerk:2009pn} without doing any analysis on the CFT data. This simplification is mainly due to the simple functional form of conformal blocks in Mellin space. It is also instructive to take one step backward and see why exchange diagrams are excluded in our solutions. Consider a scalar exchange diagram with a twist $\tau$ operator in t-channel. This produces poles at $t=\tau+2m$ in the Mellin amplitude. However, by expanding such poles in the s-channel expansion \eqref{eq:mexp1}, we have  
\begin{align}
    \frac{1}{t-\tau-2m} \sim -\frac{1}{\tau+2m}\left(1 + \frac{t}{\tau+2m} + \frac{t^2}{(\tau+2m)^2} + \frac{t^3}{(\tau+2m)^3} + \cdots\right),
\end{align}
which results in double-trace operators with arbitrarily high spin, violating the requirement $\ell\leq L$. In fact, this reflects a famous result from large spin analytic bootstrap, which states that a conformal block generally corresponds to an infinite tower of double-trace operators with unbounded spin in the cross channel \cite{Fitzpatrick:2012yx, Komargodski:2012ek}.  

Let us illustrate the first few solutions to the spin-truncated crossing equations:
\begin{align}
    \text{For } L=0,\quad \mathcal{M}(s,t) =&\ a_0.  \\
    \text{For } L=1,\quad \mathcal{M}(s,t) =&\ b_0. \\
    \text{For } L=2,\quad  \mathcal{M}(s,t) =&\ c_0 + c_1(s^2+t^2+u^2)+c_2 stu.
\end{align}
Note that solutions for $L=0$ and $L=1$ are the same because $s+t+u \equiv 4\Delta$ is constant and provides no new solution at $L=1$. More generally, the solutions for odd $L$ always match those for $L-1$. This is because we are considering the correlators of identical scalar operators, and there are no odd-spin operators in the OPE. For higher $L$, all higher degree symmetric polynomials can be expressed as a polynomial of $s^2+t^2+u^2$ and $stu$. Since the number of independent degree-$\ell$ polynomials is $\frac{\ell}{2} + 1$, we have in total $\frac{1}{2}(\frac{L}{2}+1)(\frac{L}{2}+2)$ independent solutions for $\ell\leq L$, which exactly matches the counting of independent bulk 4-point interactions modulo equations of motion and integration by parts.

\subsection{Dispersion relations and anomalous dimensions} \label{dispersionr}

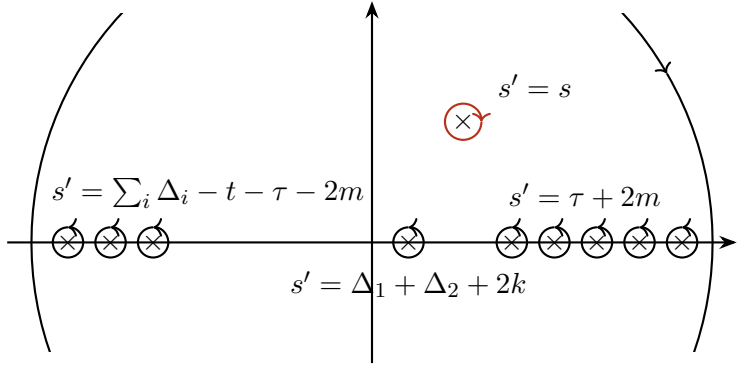
\begin{figure}
    \centering
    \begin{tikzpicture}[scale=0.8]
        \draw[thick, arrows = {-Stealth[reversed, reversed]}] (-6,0) -- (6,0);
        \draw[thick, arrows = {-Stealth[reversed, reversed]}] (0,-2) -- (0,4);
        \draw[->, thick, BrickRed] (1.5,2) + (0.3,0)  arc (360:0:0.3);
        \node at (1.5,2) {$\times$};
        \node[anchor=west] at (1.9,2.6) {$s'=s$};
        \foreach \pos in {0,...,4} {
        \draw[->, thick] (2.3+0.7*\pos,0) + (0,0.25)  arc (90:360+90:0.25);
        \node at (2.3+0.7*\pos,0) {$\times$}; 
        }
        \node[anchor=south] at (3.5,0.4) {$s'=\tau+2m$};
        \foreach \pos in {0,1,2} {
        \draw[->, thick] (-3.6-0.7*\pos,0) + (0,0.25)  arc (90:360+90:0.25);
        \node at (-3.6-0.7*\pos,0) {$\times$}; 
        }
        \node[anchor=south] at (-2.7,0.4) {$s' = \sum_i\Delta_i - t - \tau - 2m$};
        \draw[->, thick] (0.6,0) + (0,0.25)  arc (90:360+90:0.25);
        \node at (0.6,0) {$\times$}; 
        \node[anchor=north] at (0.6,-0.3) {$s' = \Delta_1+\Delta_2+2k$};
        \clip (-6,-1.8) rectangle (6,3.8);
        \draw[->, thick] (0,0) + (4.85,2.8) arc (360+30:30:5.6); 
    \end{tikzpicture}
    \caption{The contour deformation we perform in the dispersion relation $\mathcal{I}(s,t)$. The red contour is the original contour $\mathcal{C}$, while the black contour is the deformed contour. }
    \label{fig2}
\end{figure}

We now develop a method to extract anomalous dimensions from Mellin amplitudes. In Mellin space, the CFT data $a_{\tau,\ell}$ appears in pole residues as \eqref{eq:mexp} shows. To systematically extract these residues, we consider a dispersion relation. The idea  is to introduce a pole at a generic point in $s'$ and write ${\mathcal{M}}(s,t)$ as the residue of this pole using Cauchy's theorem:
\begin{align}
    \mathcal{I}(s,t) = \oint_{\mathcal{C}} \frac{\dd s'}{2\pi i} \frac{\mathcal{M}(s',t)}{(s'-s)^{N+1} (s' - (\Delta_1 + \Delta_2 + 2k))},
\end{align}
where the contour $\mathcal{C}$ is a small clockwise circle enclosing the pole at $s'=s$. This is a single variable dispersion relation in $s$, which means that we treat $t$ as constant and $u$ depends on $s$ via $s + t + u = \sum_i \Delta_i$. $N$ and $k$ are arbitrarily chosen integers, with $N$ ensuring the integrand falls off quickly enough at infinity, and $k$ serving a purpose that will become clear shortly. The integral $\mathcal{I}$ can be evaluated in two ways. On the one hand, we can evaluate the integral at $s'=s$, which gives
\begin{align}\label{eq:dis1}
    \mathcal{I}(s,t) = -\frac{1}{N!}\partial_s^N\left(\frac{\mathcal{M}(s,t)}{s-(\Delta_1+\Delta_2+2k)}\right).
\end{align}
On the other hand, we can also deform the contour to infinity, and pick up the residues of all other poles in the integrand (see Figure \ref{fig2}). 
For $N$ large enough we can neglect the arc at infinity, and the remaining contribution comes from poles in $\mathcal{M}(s',t)$ and from the pole at $s'=\Delta_1+\Delta_2+2k$  in the integrand. Due to Polyakov condition, the residue at $s'=\Delta_1+\Delta_2+2k$ vanishes, leaving only the poles from $\mathcal{M}$. These include $s$-channel poles of the form \eqref{eq:mexp}, and $u$-channel poles at $u = \tau + 2m$ (i.e., $s' = \sum_i\Delta_i - t - \tau - 2m$) since the correlator can also be expanded in terms of u-channel conformal blocks. Taking both into account, we have
\begin{align}\label{eq:dis2}
    \mathcal{I}(s,t) = \sum_{\tau,\ell}\sum_{m=0}^{\infty} \frac{a_{\tau,\ell} \mathcal{Q}_{\tau,\ell,m}(t)}{(\tau+2m-s)^{N+1}(\tau+2m-(\Delta_1+\Delta_2+2k))} + (\text{$u$ poles}).
\end{align}

We now assume that the CFT data has a perturbative expansion in a central charge $c$ \footnote{The Mellin amplitude vanishes for generalized free theory \cite{Penedones:2019tng}, so the expansion of $\mathcal{M}$ starts at $\mathcal{M}^{(1)}$.}
\begin{align}
    \mathcal{M} &= \frac{1}{c}\mathcal{M}^{(1)} + \frac{1}{c^2}\mathcal{M}^{(2)} + \cdots, \\
    \tau &= \Delta_1 + \Delta_2 + 2n + \frac{1}{c}\gamma^{(1)}_{n,\ell} + \cdots, \\
    a_{\tau,\ell} &= a^{(0)}_{n,\ell} + \frac{1}{c} a^{(1)}_{n,\ell} + \frac{1}{c^2} a^{(2)}_{n,\ell} + \cdots,
\end{align}
and expand both \eqref{eq:dis1} and \eqref{eq:dis2} to first order. We assume $\Delta_1 + \Delta_2 - \Delta_3 - \Delta_4 \in 2\mathbb{Z}$, which will be sufficient for our purposes. For \eqref{eq:dis1} we simply have 
\begin{align}
    \mathcal{I}^{(1)}(s,t) = \frac{1}{N!} \partial_s^N \left( \frac{\mathcal{M}^{(1)}(s,t)}{\Delta_1 + \Delta_2 + 2k - s} \right),
\end{align}
while \eqref{eq:dis2} expands to
\begin{align}
    \mathcal{I}(s,t) = \sum_{n,\ell} \sum_{m=0}^{\infty} \frac{\left( a^{(0)}_{n,\ell} + a^{(1)}_{n,\ell} + \cdots \right) \left( \frac{1}{2} \left( \gamma_{n,\ell}^{(1)} \right)^2 \partial_{\tau_0}^2 \mathcal{Q}_{\tau_0,\ell,m} + \cdots \right)}{\left( \tau_0 + 2m - s + \gamma_{n,\ell}^{(1)} + \cdots \right)^{N+1} \left( 2(n + m - k) + \gamma_{n,\ell}^{(1)} + \cdots \right)} + (\text{$u$ poles}),
\end{align}
where $\tau_0 = \Delta_1 + \Delta_2 + 2n$. Notably, the first two terms in the expansion of $\mathcal{Q}$ are absent due to its vanishing property at double-trace twists. At first order, $s$ pole contributions are non-zero only for $n + m = k$, and $u$ pole contributions vanish, which gives
\begin{align}
    \mathcal{I}^{(1)}(s,t) = \sum_{\ell} \sum_{n=0}^k \frac{\langle a^{(0)} \gamma^{(1)} \rangle_{n,\ell} \; \partial_{\tau_0}^2 \mathcal{Q}_{\tau_0,\ell,k-n} }{2 \left( \Delta_1 + \Delta_2 + 2k - s \right)^{N+1}},
\end{align}
where $\langle \cdots \rangle$ denotes a sum over degenerate operators which we refer to as averaged CFT data. This produces a sum rule involving double-trace operators with $\tau_0 = \Delta_1 + \Delta_2 + 2n$ for $n \leq k$:
\begin{align}\label{eq:dispersion}
    \frac{1}{N!} \partial_s^N \left( \frac{\mathcal{M}^{(1)}(s,t)}{\Delta_1 + \Delta_2 + 2k - s} \right) = \sum_{\ell} \sum_{n=0}^{k}  \frac{\langle a^{(0)} \gamma^{(1)} \rangle_{n,\ell} \; \partial_{\tau_0}^2 \mathcal{Q}_{\tau_0,\ell,k-n}(t) }{2 (\Delta_1 + \Delta_2 + 2k - s)^{N+1}} .
\end{align}
The parameter $k$ determines which double-trace operators appear in the sum rule. By taking $k = 0,1,2,\ldots$, one can recursively solve for the averaged anomalous dimensions to arbitrarily large $n$. A working example of \eqref{eq:dispersion} is provided in Section \ref{spin0}.

\section{Boundary perspective} \label{boundary}

In this section we will consider an abstract CFT consisting of scalar operators with $\Delta_+=2$ and $\Delta_-=1$. These operators are related by a shadow transformation and are dual to conformally coupled scalars in the bulk. Our goal will be to present a simple set of constraints on the boundary correlators which reproduce the shadow action in EAdS$_4$ corresponding to the scalar effective action in \eqref{shadowaction}-\eqref{shadowpot2}. 

We will start by truncating the sum over spin in the conformal block expansion of 4-point correlators and imposing crossing symmetry. The number of solutions will be greater than the number of independent 4-point vertices in the shadow action, but we will obtain the correct counting of solutions by imposing two further constraints on the anomalous dimensions. In Section \ref{bulk}, we will reproduce the results of this section from a bulk perspective, notably we will compute Witten diagrams of the bulk effective action and show that they agree with the correlators derived more abstractly from a boundary perspective in this section.  

\subsection{Solutions to crossing} \label{mellincrossing}

\begin{figure}
    \centering
    \begin{tikzpicture}[scale=0.8]
        \draw[thick, arrows = {-Stealth[reversed, reversed]}] (-0.5,0) -- (6,0);
        \draw[thick, arrows = {-Stealth[reversed, reversed]}] (0,-0.5) -- (0,6); 
        \draw[dashed] (0,5) + (-0.5,0.5) -- (5.5,-0.5) (-0.5,2.5) -- (6,2.5);
        \foreach \x in {0,...,5} \foreach \y in {0,...,5} \draw[gray, thick, fill=white] (\x,\y) circle [radius=0.09];
        \foreach \y in {0,1,2} {\pgfmathtruncatemacro{\maxx}{5-\y} \foreach \x in {0,...,\maxx} \draw[BrickRed, very thick, fill] (\x,\y) circle [radius=0.09];}
        \node[anchor=north] at (6,0) {$m$};
        \node[anchor=east] at (0,6) {$n$}; 
        \node[anchor=east] at (-0.1,2) {$\lfloor \frac{L}{2} \rfloor$}; 
        \node[anchor=east] at (-0.1,4.8) {$L$}; 
    \end{tikzpicture}
    \caption{Independent solutions for $\mathcal{M}_{++--}$, equivalent to all monomials $s^m(tu)^n$ with degree$\ \leq L$ which satisfy $m+n\leq L$ and $2n\leq L$ for $L=5$. }
    \label{fig3}
\end{figure}
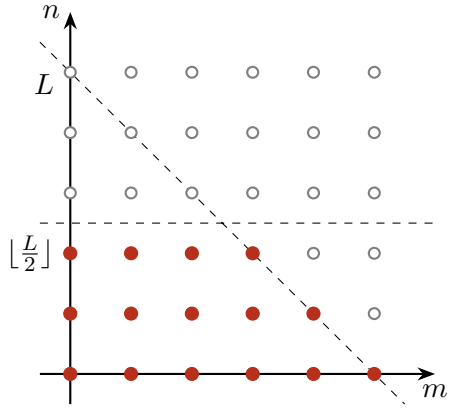

In this section, we will analyze the crossing equation for mixed correlators in Mellin space and show that after truncating the sum over spin in the conformal block expansion, only polynomial solutions are allowed. Unlike the bootstrap problem for a single scalar in EAdS described earlier, we now have three distinct types of correlators (or equivalently, Mellin amplitudes):
\begin{align}
    \mathcal{M}_{----},\quad \mathcal{M}_{++++},\quad \mathcal{M}_{++--}.
\end{align}
For the first two amplitudes with identical external dimensions, the crossing equations are the same as \eqref{eq:crossingid}. This leads to two different sets of solutions 
\begin{align}
    \mathcal{M}_{----} = \text{(degree $\leq L$ symmetric polynomial in $s,t$ and $u\equiv 4-s-t$)}, \\
    \mathcal{M}_{++++} = \text{(degree $\leq L$ symmetric polynomial in $s,t$ and $u\equiv 8-s-t$)},
\end{align}
and the total number of independent solutions for $\mathcal{M}_{----}$ and $\mathcal{M}_{++++}$ is \footnote{When $L$ is even, this reduces to $2 \times \frac{1}{2}(\frac{L}{2}+1)(\frac{L}{2}+2)$.}
\begin{align}
    2 \times \frac{ \left( \left\lfloor \frac{L}{2} \right\rfloor + 1 \right) \left( \left\lfloor \frac{L}{2} \right\rfloor + 2 \right) }{2}.
\end{align}
For $\mathcal{M}_{++--}$, the spin truncation condition in the $s$- and $t$-channels still forces $\mathcal{M}_{++--}$ to be a polynomial with degree $\leq L$, but with only $1\leftrightarrow 2$ crossing symmetry:
\begin{align}
    \mathcal{M}_{++--}(s,t) = \mathcal{M}_{++--}(s,u=6-s-t),
    \label{crossing}
\end{align}
where we noted that $2(\Delta_-+\Delta_+)=6$. The resulting solutions are therefore
\begin{align}
    \mathcal{M}_{++--} = \text{(symmetric polynomial in $t,u$ with degree $\leq L$ in $s,t$)},
\end{align}
where we replace $u$ with $6-s-t$ when counting the degree. Symmetric polynomials of $t$ and $u$ can be written in terms of the basis $t+u$ and $t u$. However $t+u=6-s\sim s$, so all the independent solutions are polynomials of $s$ and $tu$.
The first few examples are 
\begin{align}
    \text{For } L=0,\quad \mathcal{M}_{++--} =&\ a_0.  \\
    \text{For } L=1,\quad \mathcal{M}_{++--} =&\ b_0 + b_1 s. \\
    \text{For } L=2,\quad  \mathcal{M}_{++--} =&\ c_0 + c_1 s + c_2 s^2 + c_3 tu + c_4 stu.
\end{align}
Unlike the solutions for identical external operators, these solutions include odd degree terms corresponding to odd spin operators in the $t$-channel OPE $\mathcal{O}_{+}\times \mathcal{O}_{-}$. For generic $L$, the number of independent solutions for $\mathcal{M}_{++--}$ is the same as the number of monomials $s^m (tu)^n$ satisfying $m+n\leq L$ and $2n\leq L$, which is given by
\begin{align}
    \frac{(L+1)(L+2)}{2} - \frac{ \left(L - \left\lfloor \frac{L}{2} \right\rfloor \right) \left(L - \left\lfloor \frac{L}{2} \right\rfloor + 1 \right) }{2}.
\end{align} 
See Figure \ref{fig3} for an illustration. Together with the solutions for $\mathcal{M}_{++++}$ and $\mathcal{M}_{----}$, the total number of independent crossing-symmetric solutions is 
\begin{align}
    \frac{ \left( \left\lfloor \frac{L}{2} \right\rfloor + 1 \right) \left( \left\lfloor \frac{L}{2} \right\rfloor + 2L + 6 \right) }{2}.
\label{totalsolutions}
\end{align}
which is greater then the number of free coefficients in the shadow action in \eqref{shadowpot2} up to spin $L$.

Although crossing symmetry provides essential constraints, it alone cannot fully determine the shadow action in \eqref{shadowaction}-\eqref{shadowpot2}. Indeed, since this action was derived from an EFT of a single scalar field in dS$_4$, it does not contain odd-spin interactions and has fewer free coefficients than the number of solutions to the truncated crossing equations in \eqref{totalsolutions}. In order to derive this action from a purely CFT perspective, we will therefore impose additional constraints which we describe below.

\subsection{Anomalous dimension constraints}\label{sec:adcon}

To further constrain the solutions to the crossing equations, we will impose two simple constraints on the anomalous dimensions extracted from these Mellin amplitudes using the dispersion relations derived in Section \ref{dispersionr}.

First note that the operators appearing in the OPE of these amplitudes are double-trace operators taking the schematic form 
\begin{align}
    :\mathcal{O}_- \square^n \partial^\ell \mathcal{O}_-:,\quad :\mathcal{O}_+ \square^n \partial^\ell \mathcal{O}_+:,\quad :\mathcal{O}_+ \square^n \partial^\ell \mathcal{O}_-:.
\end{align}
We denote these operators respectively as  
\begin{align}
    [\mathcal{O}_-\mathcal{O}_-]_{n,\ell},\quad [\mathcal{O}_+\mathcal{O}_+]_{n,\ell},\quad [\mathcal{O}_+\mathcal{O}_-]_{n,\ell}. 
\end{align}
These operators have spin $\ell$ and classical twist $2+2n$, $4+2n$ and $3+2n$, and acquire anomalous dimensions due to interactions. Operators $[\mathcal{O}_-\mathcal{O}_-]_{n+1,\ell}$ and $ [\mathcal{O}_+\mathcal{O}_+]_{n,\ell}$ share the same classical quantum numbers. As a consequence, they are degenerate in the OPE and mix together. After taking into account quantum corrections, the eigenstates of the dilatation operator are linear combinations of these two operators, which we denote by
\begin{align}
    [\mathcal{O}\mathcal{O}]_{n,\ell}^+ =&\ A_{1} [\mathcal{O}_-\mathcal{O}_-]_{n+1,\ell} + A_{2} [\mathcal{O}_+\mathcal{O}_+]_{n,\ell},\label{eq:degop1}\\
    [\mathcal{O}\mathcal{O}]_{n,\ell}^- =&\ A_{3} [\mathcal{O}_-\mathcal{O}_-]_{n+1,\ell} + A_{4} [\mathcal{O}_+\mathcal{O}_+]_{n,\ell}\label{eq:degop2}.
\end{align}
We call the anomalous dimensions of these eigenstates $\gamma^{\text{mixed},\pm}_{n,\ell}$. In contrast, there is no mixing issue for the operator $[\mathcal{O}_+\mathcal{O}_-]_{n,\ell}$ since it is the only double-trace operator appearing in the t-channel OPE of $\mathcal{M}_{++--}$, so we call its anomalous dimension $\gamma^{\text{pure}}_{n,\ell}$.

To determine $\gamma^{\text{mixed},\pm}_{n,\ell}$ we need to solve a set of linear equations involving $\langle (a^{(0)})_{n,\ell} \rangle$ and $\langle (a^{(0)}\gamma^{(1)})_{n,\ell} \rangle$. On the other hand, we can directly obtain $\gamma^{\text{mixed},\pm}$ as the eigenvalues of the following matrix:
\begin{align} \label{eq:mat}
    M_{n,\ell} =  \begin{pmatrix} 
          \frac{\langle (a^{(0)}\gamma^{(1)})_{n+1,\ell} \rangle_{----}}{\langle a^{(0)}_{n+1,\ell} \rangle_{----}} &  \frac{\langle (a^{(0)}\gamma^{(1)})_{n,\ell} \rangle_{++--}}{\sqrt{\langle a^{(0)}_{n,\ell} \rangle_{++++} \langle a^{(0)}_{n+1,\ell} \rangle_{----} }} \\[5mm] 
        \frac{\langle (a^{(0)}\gamma^{(1)})_{n,\ell} \rangle_{++--}}{\sqrt{\langle a^{(0)}_{n,\ell} \rangle_{++++} \langle a^{(0)}_{n+1,\ell} \rangle_{----} }} & \frac{\langle (a^{(0)}\gamma^{(1)})_{n,\ell} \rangle_{++++}}{\langle a^{(0)}_{n,\ell} \rangle_{++++}} 
    \end{pmatrix},
\end{align}
which we derive in Appendix \ref{unmixing} using techniques first developed in the context of $\mathcal{N}=4$ SYM \cite{Aprile:2017xsp}. The averaged CFT data $\langle a^{(0)}\gamma^{(1)} \rangle$ can be obtained from 4-point Mellin amplitudes using the dispersion relation \eqref{eq:dispersion}, while the free OPE coefficients $\langle a^{(0)} \rangle$ are given by \cite{Fitzpatrick:2011dm} 
\begin{align}
\label{opecoeffs}
    \langle a^{(0)}_{n,\ell} \rangle_{ss's's} =&\ \frac{(-)^{ss'}\, (1+(-1)^{\ell}\delta_{ss'})(\Delta_{s}-\frac{d}{2}+1)_n(\Delta_{s'}-\frac{d}{2}+1)_n }{\ell!n!(\ell+\frac{d}{2})_n(\Delta_{s}+\Delta_{s'}-d+n+1)_n(\Delta_{s}+\Delta_{s'}-d+\ell+n)_n} \nn
    &\times \frac{(\Delta_{s})_{\ell+n}(\Delta_{s'})_{\ell+n}}{(\Delta_{s}+\Delta_{s'}+\ell+2n-1)_\ell},
\end{align}
where $s,s'=\pm$ and $(a)_b = \frac{\Gamma(a+b)}{\Gamma(a)}$. There is also an extra factor $(-)^{ss'}$ because the two-point function of $\mathcal{O}_{+}$ has a minus sign
\begin{align}
    \langle \mathcal{O}_{+}(x_1)\mathcal{O}_{+}(x_2) \rangle = -\frac{1}{\left(x_{12}^2\right)^2},
\label{2ptplus}
\end{align}
which originates from the sign of the kinetic term of $\phi_+$ in the shadow action \eqref{shadowaction}. Given this CFT data, we can then compute $\gamma^{\text{mixed},\pm}$:
\begin{align}\label{eq:gammamix}
    M_{n,\ell} \equiv \begin{pmatrix}
        M^{--}_{n,\ell} & M^{+-}_{n,\ell} \\ M^{+-}_{n,\ell} & M^{++}_{n,\ell}
    \end{pmatrix} \implies \gamma^{\text{mixed},\pm}_{n,\ell} = \frac{M^{++}_{n,\ell} + M^{--}_{n,\ell} \pm \sqrt{(M^{++}_{n,\ell}-M^{--}_{n,\ell})^2+4 (M^{+-}_{n,\ell})^2}}{2}.
\end{align}

After diagonalising the matrix in \eqref{eq:mat} we then impose the following additional constraints \footnote{We can also take  $\gamma^{\text{mixed},-}_{n,\ell} = \gamma^{\text{pure}}_{n+\frac{1}{2},\ell}$ and $\gamma^{\text{mixed},+}_{n,\ell} = 0$ in the constraint. The solutions would not change. }:
\begin{align}\label{eq:ancon}
    \gamma^{\text{mixed},-}_{n,\ell} = 0\,, \qquad 
    \gamma^{\text{mixed},+}_{n,\ell} = \gamma^{\text{pure}}_{n+\frac{1}{2},\ell}\,.
\end{align}
Remarkably, these constraints in combination with crossing symmetry can be used to reconstruct the shadow action in $\mathrm{EAdS}_4$. They reflect that there is only one underlying scalar field $\phi$ in the original de Sitter EFT.  

One might wonder if we can impose a different set of constraints: 
\begin{equation}
    \gamma^{\text{mixed},-}_{n,\ell} = \gamma^{\text{mixed},+}_{n,\ell} = \gamma^{\text{pure}}_{n+\frac{1}{2},\ell}\,,
\end{equation}
which also seem natural. However, these constraints indicate that the unmixing matrix \eqref{eq:mat} has two identical eigenvalues and must therefore be proportional to the identity matrix. This forces the off-diagonal element $M^{+-}_{n,\ell}$ (and thus the whole amplitude $\mathcal{M}_{++--}$) to vanish and implies that there is no interaction between $\phi_+$ and $\phi_-$, leading to two decouple CFTs, which is clearly not the case for the shadow action. 

We present detailed bootstrap computations for spin $L\leq 2$ below. These bootstrap results will be compared with the direct bulk computations using the shadow action in Section \ref{bulk}.

\subsection{Spin-0} \label{spin0}

We begin with a detailed illustration of the bootstrap procedure for the simplest case, notably spin-0. The solutions of crossing equations \eqref{eq:crossingid} and \eqref{crossing} in this case are simply
\begin{align}
    \mathcal{M}_{----} =&\ a_0,\\ 
    \mathcal{M}_{++++} =&\ b_0,\\ 
    \mathcal{M}_{++--} =&\ c_0.
\end{align}
We now use the dispersion relation \eqref{eq:dispersion} to recursively solve for the averaged CFT data $\langle (a^{(0)} \gamma^{(1)})_{n,\ell} \rangle$. Take $\mathcal{M}_{----}$ as an example. Taking $N=0$ in the dispersion relation \eqref{eq:dispersion} and truncating the sum over spin at $L=0$ then gives
\begin{align}
    \frac{a_0}{2 + 2k - s} = \sum_{\ell=0}^0 \sum_{n=0}^{k}  \frac{\langle (a^{(0)} \gamma^{(1)})_{n,\ell} \rangle_{----} \; \partial_{\tau_0}^2 \mathcal{Q}_{\tau_0,\ell,k-n}(t) }{2 (2 + 2k - s)} .
\end{align}
Taking $k=0,1,2,...$ implies the following equations:
\begin{align}
    \frac{a_0}{2-s} =&\ \frac{\langle (a^{(0)} \gamma^{(1)})_{0,0} \rangle_{----}}{4(2-s)}, \\ 
    \frac{a_0}{4-s} =&\ \frac{\langle (a^{(0)} \gamma^{(1)})_{0,0} \rangle_{----}}{6(4-s)} + \frac{3\langle (a^{(0)} \gamma^{(1)})_{1,0} \rangle_{----}}{2(4-s)}, \\ 
    \frac{a_0}{6-s} =&\ \frac{2\langle (a^{(0)} \gamma^{(1)})_{0,0} \rangle_{----}}{15(6-s)} + \frac{12\langle (a^{(0)} \gamma^{(1)})_{1,0} \rangle_{----}}{7(6-s)} + \frac{15\langle (a^{(0)} \gamma^{(1)})_{2,0} \rangle_{----}}{2(6-s)}, \\
     \vdots\, & \nonumber
\end{align}
These equations can be solved recursively giving 
\begin{align}
    \langle (a^{(0)} \gamma^{(1)})_{n,0} \rangle_{----} = \frac{\sqrt{\pi }\, 4^{-2 n + 1} \Gamma (2 n+1)}{(2 n+1) \Gamma \left(2 n+\frac{1}{2}\right)}\, a_0.
\end{align}
Similarly for $\mathcal{M}_{++++}$ and $\mathcal{M}_{++--}$ we obtain
\begin{align}
    \langle (a^{(0)} \gamma^{(1)})_{n,0} \rangle_{++++} =&\ \frac{\sqrt{\pi }\, 4^{-2 n-1} (n+1)^2 \Gamma (2 n+4)}{3 \Gamma \left(2 n+\frac{5}{2}\right)}\, b_0, \\ 
    \langle (a^{(0)} \gamma^{(1)})_{n,0} \rangle_{++--} =&\ \frac{\sqrt{\pi }\, 4^{-2 n-1} (n+1) \Gamma (2 n+3)}{\Gamma \left(2 n+\frac{5}{2}\right)}\, c_0.
\end{align}
According to \eqref{eq:mat} the unmixing matrix is then
\begin{align}
    M_{n,\ell} = \begin{pmatrix}
        a_0 & c_0 \\
        c_0 & \frac{b_0}{3}
    \end{pmatrix} \delta_{\ell,0},
\end{align}
which happens to be independent of $n$. The eigenvalues are 
\begin{align}
    \gamma^{\text{mixed},\pm}_{n,\ell} = \frac{3 a_0+b_0\pm\sqrt{(3a_0-b_0)^2+36 c_0^2}}{6}\, \delta_{\ell,0}. 
\end{align}

For $\gamma^{\text{pure}}$, we compute the t-channel decomposition of $\mathcal{M}_{++--}$. This can be done by interchanging $s\leftrightarrow t$ in the dispersion relation and the result is 
\begin{align}
    \langle (a^{(0)} \gamma^{(1)})_{n,0} \rangle_{+--+} =&\ \frac{\sqrt{\pi }\, 2^{-4 n-1} \Gamma (2 n+3)}{\Gamma \left(2 n+\frac{3}{2}\right)}c_0.
\end{align}
Since $[\mathcal{O}_+\mathcal{O}_-]_{n,\ell}$ is non-degenerate, its anomalous dimension is simply 
\begin{align}
    \gamma^{\text{pure}}_{n,\ell} = \frac{\langle (a^{(0)} \gamma^{(1)})_{n,0} \rangle_{+--+}}{\langle a^{(0)} _{n,0} \rangle_{+--+}}\delta_{\ell,0} = -2c_0\, \delta_{\ell,0}.
\end{align}

The anomalous dimension constraints in \eqref{eq:ancon} then imply
\begin{align}
    \frac{3 a_0+b_0 - \sqrt{(3a_0-b_0)^2+36 c_0^2}}{6} =&\ 0,\\
    \frac{3 a_0+b_0 + \sqrt{(3a_0-b_0)^2+36 c_0^2}}{6} =& -2c_0.
\end{align}
Although the equations involve complicated-looking square roots, the solution is remarkably simple: 
\begin{align}
    b_0 = 3a_0,\quad c_0=-a_0. 
\end{align}
Thus all Mellin amplitudes are determined by a single parameter
\begin{align}
    \mathcal{M}_{----} =&\ a_0,\label{mellinspin01}\\ 
    \mathcal{M}_{++++} =&\ 3a_0,\\ 
    \mathcal{M}_{++--} =&-a_0.\label{mellinspin03}
\end{align}
In Section \ref{bulk}, we will show that these Mellin amplitudes arise from the following bulk interaction vertex:
\begin{align}
    \frac{\pi^2}{3}a_0 (\phi_+^4-6\phi_+^2\phi_-^2+\phi_-^4),
\end{align}
which exactly reproduces the spin-0 part of shadow action \eqref{shadowpot2} coming from a $\phi^4$ interaction in dS$_4$. 

\subsection{Spin-1} \label{spin1_bootstrap}

Next we consider solutions to the crossing equations with spin truncated at $L=1$. The most general crossing-symmetric solutions are
\begin{align}
    \mathcal{M}_{----} =&\ a_0, \\
    \mathcal{M}_{++++} =&\ b_0 , \\
    \mathcal{M}_{++--} =&\ c_0 + c_1 s. 
\end{align}
The CFT data for $\mathcal{M}_{----}$ and $\mathcal{M}_{++++}$ remain identical to the spin-0 case, while applying \eqref{eq:dispersion} to $\mathcal{M}_{++--}$ yields
\begin{align} 
    \langle (a^{(0)} \gamma^{(1)})_{n,0} \rangle_{++--} =&\ \frac{\sqrt{\pi }\, 4^{-2 n-1} (n+1) \Gamma (2 n+3)}{\Gamma \left(2 n+\frac{5}{2}\right)} \left(c_0 + \frac{2}{3}\left(2 n^2+5n+6\right)c_1 \right), \\
    \langle (a^{(0)} \gamma^{(1)})_{n,0} \rangle_{+--+} =&\ \frac{\sqrt{\pi }\, 2^{-4 n-1} \Gamma (2 n+3)}{\Gamma \left(2 n+\frac{3}{2}\right)}\, \left(c_0 - \frac{1}{3} \left(2 n^2+3n-4\right) c_1\right), \label{spin1data0} \\
    \langle (a^{(0)} \gamma^{(1)})_{n,1} \rangle_{+--+} =&\ -\frac{\sqrt{\pi }\, 2^{-4 n-3} (n+1)^2 \Gamma (2 n+5)}{3 (2 n+3) \Gamma \left(2 n+\frac{5}{2}\right)} c_1.
\label{spin1data}
\end{align}
According to \eqref{eq:mat}, the unmixing matrix is 
\begin{align}
    M_{n,\ell} = \begin{pmatrix}
        a_0 & c_0 + \frac{2}{3}\left(2 n^2+5n+6\right)c_1 \\
        c_0 + \frac{2}{3}\left(2 n^2+5n+6\right)c_1 & \frac{b_0}{3}
    \end{pmatrix} \delta_{\ell,0},
\end{align}
from which we can easily obtain $\gamma^{\text{mixed},\pm}$ using \eqref{eq:gammamix}. Moreover, from \eqref{spin1data0} and \eqref{spin1data} we have 
\begin{align}
    (\gamma^{\text{pure}})_{n,\ell} = \left(-2c_0 + \frac{2}{3}  \left(2 n^2+3n-4\right) c_1\right)\delta_{\ell,0} + \frac{2}{9} (n+1) (2 n+3)c_1\, \delta_{\ell,1} . 
\end{align}
The anomalous dimension constraints in \eqref{eq:ancon} for $\ell=1$ require $(\gamma^{\text{pure}})_{n,1}=0$, which immediately forces $c_1=0$. The system then reduces to the spin-0 solution
\begin{align}
    b_0 = 3a_0,\quad c_0=-a_0, \quad c_1=0,
\end{align}
so the amplitudes are the same as the spin-0 case. In Section \ref{bulk}, we will show that this is consistent with bulk expectations, since the underlying EFT in de Sitter space does not contain odd-spin interactions.

\subsection{Spin-2} \label{spin2bootstrap}

Finally, let us consider the spin-2 truncation of the crossing equations. As we will see in Section \ref{bulk}, this corresponds to four and six-derivative interactions in the bulk. 

The crossing symmetric solutions for $L=2$ are 
\begin{align}
    \mathcal{M}_{----} =&\ a_0 + a_1(s^2+t^2+u^2)+a_2 stu, \\ 
    \mathcal{M}_{++++} =&\ b_0 + b_1(s^2+t^2+u^2)+b_2 stu, \\ 
    \mathcal{M}_{++--} =&\ c_0 + c_1 s + c_2 s^2 + c_3 tu + c_4 stu.
\end{align}
Using the dispersion relation we can work out the unmixed anomalous dimensions. Since their expressions are lengthy we record them in Appendix \ref{spin2}. Imposing the the anomalous dimension constraints in \eqref{eq:ancon} then fixes the coefficients as follows: 
\begin{gather}
    b_0=3 a_0-680 a_1-2144 a_2,\quad b_1=35 a_1+42 a_2,\quad b_2=63 a_2, \\
    c_0=-a_0-\frac{374 }{3}a_1+\frac{4 }{15}a_2 ,\quad c_1=58 a_1+\frac{151 }{5}a_2,  \\
    c_2=-10 a_1-5 a_2,\quad c_3=10 a_1+4 a_2,\quad c_4=-7 a_2,
\end{gather}
and the resulting Mellin amplitudes are 
\begin{align}
    \mathcal{M}_{----} =&\ a_0 + a_1(s^2+t^2+u^2)+a_2 stu,
    \nonumber
    \\ 
    \mathcal{M}_{++++} =&\ 3 a_0 + a_1 \left(-680 \!+\! 35 (s^2 \!+\! t^2 \!+\! u^2)\right)\! +\! a_2 \left(-2144 \!+\! 42(s^2 \!+\! t^2 \!+\! u^2) \!+\! 63 s t u\right), 
    \nonumber
    \\ 
    \mathcal{M}_{++--} =& -\!a_0 \!+\! a_1 \!\left(\!-\frac{374}{3}\!+\!58 s \!-\!10 s^2\!+\!10 t u\!\right)\!+\!a_2\! \left(\!\frac{4}{15}\!+\!\frac{151}{5}s \!-\!5 s^2 \!+\!4 t u \!-\!7 s t u\!\right)\!.
    \label{spin2boot}
\end{align}
The $a_0$ solution corresponds to the spin-0 case, while $a_1$ and $a_2$ corresponds to two- and four-derivative interactions in the bulk, respectively.

\section{Bulk perspective} \label{bulk}

In this section we will consider a general local EFT of two conformally coupled scalar fields $\phi_{\pm}$ dual to boundary operators with $\Delta_{+}=2$ and $\Delta_{-}=1$. First we show that the 4-point interaction vertices are in one-to-one correspondence with the solutions of the truncated crossing equations. After computing 4-point Witten diagrams and matching them with the Mellin amplitudes bootstrapped in the previous section, we will then show that the anomalous dimension constraints proposed in the previous section fix the coefficients of the EFT to be precisely those of the shadow action in \eqref{shadowaction}-\eqref{shadowpot2}. 

Our starting point will be the following EFT in EAdS$_4$:
\begin{align}
S_{EAdS}&=\int \frac{dZ d^3x}{Z^4}  \bigg(-(\partial\f_+)^2 +m^2 \f^2_+ +(\partial\f_-)^2 -m^2 \f^2_-+V(\f_+, \f_-)\bigg)\,,
\label{bulkgeneral}
\end{align}
where $m^2=-2$, corresponding to conformally coupled scalars, and the metric is given by \eqref{adsmetric}. Note that covariant derivatives are contracted with a flat Euclidean metric. Up to six derivatives, the potential is given by
\begin{align}
V(\f_+, \f_-)=V_0(\f_+, \f_-)+V_2(\f_+, \f_-)+V_4(\f_+, \f_-)+V_6(\f_+, \f_-)+\cdots,
\end{align}
where
\begin{align}
V_0(\f_+, \f_-) &= \lambda^{(0)}_{1} {\f^4_+} +\lambda^{(0)}_{2} \f^4_- + \lambda^{(0)}_{3} \f^2_+\f^2_-\,, \label{0der}\\
V_2(\f_+, \f_-) &= \lambda^{(2)}_{1} \f_+ \f_-\partial_\mu \f_+ \partial^\mu \f_-\,, \label{2derindep}\\
V_4(\f_+, \f_-) &= \lambda^{(4)}_{1}(\partial \f_+)^4+ \lambda^{(4)}_{2}(\partial \f_-)^4+\lambda^{(4)}_{3}(\partial \f_+)^2 (\partial \f_-)^2+\lambda^{(4)}_{4}(\partial_{\mu} \f_+ \partial^{\mu} \f_-)^2\,, \label{4derindep}\\
V_6(\f_+, \f_-) &= \lambda^{(6)}_{1}(\partial \phi_+)^2 (\nabla_\mu \nabla_\nu  \phi_{+})^2+\lambda^{(6)}_{2}(\partial \phi_-)^2 (\nabla_\mu \nabla_\nu  \phi_{-})^2 \nn
&+\lambda_{3}^{(6)}\left[(\partial\phi_{+})^{2}(\nabla_{\mu}\nabla_{\nu}\phi_{-})^{2}+2(\partial_{\sigma}\phi_{+}\partial^{\sigma}\phi_{-})(\nabla_{\mu}\nabla_{\nu}\phi_{+})(\nabla^{\mu}\nabla^{\nu}\phi_{-})\right]
\label{6derindep}\,.
\end{align}
We assume that the two scalars have kinetic terms with opposite signs and that the potential has a $\mathbb{Z}_2$ symmetry $\f_{\pm}\rightarrow -\f_{\pm}$, but otherwise the action contains the most general local interaction vertices modulo equations of motion and integration by parts. 

Note that the two interaction vertices in the second line of \eqref{6derindep} have the same coupling. This can be understood by considering what kind of amplitudes they give in flat space. In particular, if we use them to compute a 4-point tree-level amplitude with legs $1,2$ corresponding to $+$ and legs $3,4$ corresponding to $-$, then the first term gives $s^3$ and the second term gives $t^3+u^3$. While each vertex individually corresponds to a spin-3 interaction, if we add them together, we get $s^3+t^3+u^3$ which can be written as a linear combination of $s^2+t^2+u^2$ and $stu$ after using the constraint $s+t+u=4m^2$, which are spin-2 interactions. We therefore only consider this particular linear combination. After doing so, we can easily see that the number of interaction vertices agrees with the counting of solutions to the truncated crossing equations in \eqref{totalsolutions}, notably there are three coefficients at spin-0, 4 coefficients up to spin-1, and 11 coefficients up to spin-2. Our goal will then be to fix the coefficients to those of \eqref{shadowpot2} by computing 4-point boundary correlators and comparing them to the bootstrap results in the previous section.

\subsection{Witten diagrams}

Using the EFT in \eqref{bulkgeneral}, we will compute tree-level Witten diagrams describing four-point correlators of the dual CFT. We will then convert them to Mellin amplitudes and compare to the bootstrap results in the previous section. In order to compute Witten diagrams, it is convenient to use the embedding space formalism \cite{Costa:2011mg}. Euclidean AdS$_4$ with unit radius is defined by the following hyperboloid embedded in $5$-dimensional Minkowski spacetime $\mathbb{R}^{4,1}$:
\begin{align}
-(X^0)^2+(X^1)^2+(X^2)^2+(X^3)^2+(X^4)^2=-1, \quad X^0 >0\,.
\end{align}
Moreover the boundary of EAdS$_4$ is defined by the space of null rays in $\mathbb{R}^{4,1}$
\begin{align}
-(P^0)^2+(P^1)^2+(P^2)^2+(P^3)^2+(P^4)^2=0\,.
\end{align}
The bulk-to-boundary propagators can then be written in embedding space coordinates as \cite{Witten:1998qj}
\begin{align}
\label{propagator}
G^{\D_\f}_{B \partial}(P,X)&= \sqrt{\frac{\Gamma(\D_\f)}{2\pi^{\frac{3}{2}}\Gamma(\D_\f-\frac{1}{2})}}\frac{1}{(-2P\cdot X)^{\D_\f}}\,,
\end{align}
where $\Delta_{\phi}$ is given by \eqref{deltamass}. Correlators can be written in terms of embedding space coordinate as
\begin{align}
\langle \f_1\f_2\f_3\f_4\rangle=\frac{1}{P_{12}^{\frac{\D_1+\D_2}{2}}P_{34}^{\frac{\D_3+\D_4}{2}}}\left(\frac{P_{14}}{P_{24}}\right)^a \left(\frac{P_{14}}{P_{13}}\right)^b \mathcal{G}_{\{\D_i\}}(z,\zb),
\end{align}
where $P_{ij}=-2P_i \cdot P_j$, $a,b,$ and ${\{\D_i\}}$ are defined below \eqref{4ptfn}, and the cross-ratios are given in terms of embedding coordinates as
\begin{align}\label{crossratio}
z \zb=\frac{P_{12}P_{34}}{P_{13}P_{24}}\,,\quad (1-z)(1-\zb)=\frac{P_{14}P_{23}}{P_{13}P_{24}}\,.
\end{align}
Note that embedding coordinates are related to Poincar\'e patch coordinates in \eqref{adsmetric} as follows:
\begin{align}
    X^0 & = \frac{1+\vec{x}^2+Z^2}{2Z}\,,\qquad
    X^{\mu} = \frac{x^\mu}{Z}\,, \qquad
    X^4 =\frac{1-\vec{x}^2-Z^2}{2Z}\,,\nn 
    P^0 &=\frac{1+\vec{x}^2}{2}\,,\qquad P^{\mu} =x^{\mu}\,,\qquad P^4=\frac{1-\vec{x}^2}{2}\,,
\end{align}
where ${x}^\mu \in \mathbb{R}^3$ and $Z>0$.

\subsection{Generalised free field theory}

\begin{figure}[h]
{\small
\begin{center}
\begin{tikzpicture}[scale=0.3]
\draw (0,0) circle (5cm);
\draw (-4,-3) -- (4,-3);
\draw (-4,3) -- (4,3);
\filldraw[black] (4,3) circle (4pt) node[anchor=west]{$P_2$};
\filldraw[black] (4,-3) circle (4pt) node[anchor=west]{$P_4$};
\filldraw[black] (-4,3) circle (4pt) node[anchor=east]{$P_1$};
\filldraw[black] (-4,-3) circle (4pt) node[anchor=east]{$P_3$};
\node at (7,0) {$+$};
\node at (21,0) {$+$};
\draw (14,0) circle (5cm);
\draw (10,-3) -- (10,3);
\draw (18,-3) -- (18,3);
\filldraw[black] (10,3) circle (4pt) node[anchor=east]{$P_1$};
\filldraw[black] (10,-3) circle (4pt) node[anchor=east]{$P_3$};
\filldraw[black] (18,3) circle (4pt) node[anchor=west]{$P_2$};
\filldraw[black] (18,-3) circle (4pt) node[anchor=west]{$P_4$};
\draw (28,0) circle (5cm);
\filldraw[black] (24,3) circle (4pt) node[anchor=east]{$P_1$};
\filldraw[black] (24,-3) circle (4pt) node[anchor=east]{$P_3$};
\filldraw[black] (32,3) circle (4pt) node[anchor=west]{$P_2$};
\filldraw[black] (32,-3) circle (4pt) node[anchor=west]{$P_4$};
\draw (24,3) -- (32,-3);
\draw (24,-3) -- (27.6,-0.1);
\draw (32,3) -- (28.4,0.1);
\draw[black] (27.6,-0.1) to[out=45,in=135] (28.4,0.1);
\end{tikzpicture}
\caption{Disconnected Witten diagrams for generalised free theory}
      \label{fig:wittendiagramdisc}
\end{center}}
\end{figure}
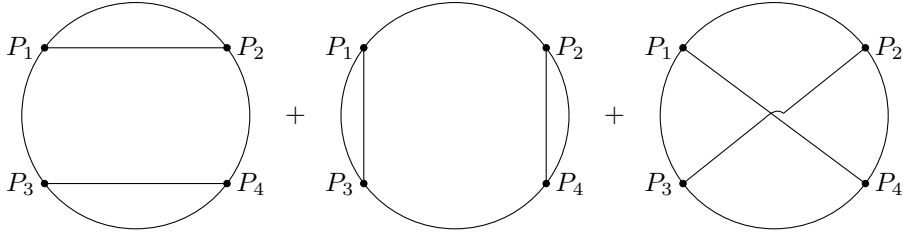

While anomalous dimensions can be conveniently extracted from from Mellin amplitudes using dispersion relations, free OPE coefficients are most easily derived by computing the conformal block expansion of generalised free field theory (GFF) correlators in position space. The latter can be computed from disconnected Witten diagrams shown in Figure \ref{fig:wittendiagramdisc}, which are simply given by products of the 2-point functions in \eqref{2ptplus} and
\begin{equation}
\left\langle \mathcal{O}_{-}(x_{1})\mathcal{O}_{-}(x_{2})\right\rangle =\frac{1}{x_{12}^{2}}.
\end{equation}

Using the definition of bulk-to-boundary propagators in \eqref{propagator} one readily finds that
\begin{align}
\langle \f_1\f_2\f_3\f_4\rangle_{GFF} &= \frac{1}{P_{12}^{\frac{\D_1+\D_2}{2}}P_{34}^{\frac{\D_3+\D_4}{2}}}\left(\frac{P_{14}}{P_{24}}\right)^a \left(\frac{P_{14}}{P_{13}}\right)^b+ \text{(permutations)}\,.
\end{align}
The GFF correlator can then be decomposed into conformal blocks \eqref{partialwaves} associated to the exchanged operators $\D_{n, \ell}=\D_1+\D_2+2n+\ell$, which  gives the GFF OPE coefficients. For our purposes $\Delta_i$ can be 1 or 2. After performing the conformal block decomposition in \eqref{partialwaves} we can then deduce the OPE coefficients in \eqref{opecoeffs}.

\subsection{Spin-0}

Now let us consider 4-point interacting diagrams using the spin-0 part of the EFT in \eqref{0der}.
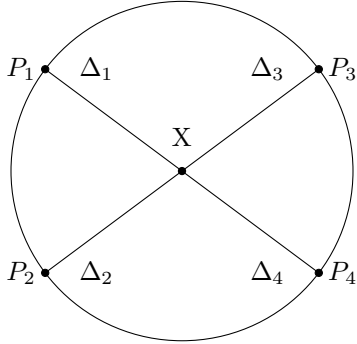
\begin{figure}[h]
{\small
\begin{center}
\begin{tikzpicture}[scale=0.45]
\draw (0,0) circle (5cm);
\draw (-4,-3) -- (4,3);
\draw (-4,3) -- (4,-3);
\filldraw[black] (0,0) circle (3pt) ;
\filldraw[black] (4,3) circle (3pt) node[anchor=west]{$P_3$};
\filldraw[black] (4,-3) circle (3pt) node[anchor=west]{$P_4$};
\filldraw[black] (-4,3) circle (3pt) node[anchor=east]{$P_1$};
\filldraw[black] (-4,-3) circle (3pt) node[anchor=east]{$P_2$};
  \node at (2.5,3) {$\D_3$};
  \node at (-2.5,3) {$\D_1$};
  \node at (2.5,-3) {$\D_4$};
  \node at (-2.5,-3) {$\D_2$};
  \node at (0,1) {X};
\end{tikzpicture}
\caption{Contact Witten diagram for quartic vertex}
      \label{fig:wittendiagram}
\end{center}}
\end{figure}
The Witten diagrams consist of four bulk-to-boundary propagators attached to an interaction vertex which is integrated over bulk and are given by $D$-functions \cite{DHoker:1999kzh, Arutyunov:2002fh, Gary:2009ae}: 
\begin{align}
D_{\Delta_1 \Delta_2 \Delta_3 \Delta_4} =  \int_{AdS} dX\   G^{\D_1}_{B\partial}(P_1,X)\ G^{\D_2}_{B\partial}(P_2,X)\ G^{\D_3}_{B\partial}(P_3,X)\ G^{\D_4}_{B\partial}(P_4,X), \label{ddef}
\end{align}
where each of the $\D_{i}$'s can take value either $1$ or $2$. We can further strip off kinematic factors and define $\db$-functions which only depend on cross-ratios:
\begin{align}
    D_{\Delta_1\Delta_2\Delta_3\Delta_4} =  \frac{1}{P_{12}^{\frac{\D_1+\D_2}{2}}P_{34}^{\frac{\D_3+\D_4}{2}}}\left(\frac{P_{14}}{P_{24}}\right)^a \left(\frac{P_{14}}{P_{13}}\right)^b  \db_{\Delta_1\Delta_2\Delta_3\Delta_4}(z,\zb)\,.
\end{align}
The Mellin amplitudes of $\db$-functions are given by constants \cite{Penedones:2010ue}:
\begin{align}\label{eq:Dmellin}
    \overbar{D}_{\Delta_1\Delta_2\Delta_3\Delta_4} =&\ \int_{-i\infty}^{i\infty} \frac{\dd s\, \dd t}{(2\pi i)^2} U^{\frac{s}{2}}V^{\frac{t-\Delta_2-\Delta_3}{2}} \frac{\pi ^{3/2}  \Gamma \left(\frac{\Delta_1+\Delta_2+\Delta_3+\Delta_4}{2}-\frac{3}{2}\right)}{2 \Gamma (\Delta_1) \Gamma (\Delta_2) \Gamma (\Delta_3) \Gamma (\Delta_4)}\, \Gamma_{\{\Delta_i\}}(s,t)\, .
\end{align}

The vertices in \eqref{0der} then give rise to the following boundary correlators: 
\begin{align}
\langle \f_{+} \f_{+} \f_{+} \f_{+}\rangle & = 24\, \lambda^{(0)}_{1}  D_{2222}\,,\label{0derdfun1}\\
\langle \f_{-} \f_{-} \f_{-} \f_{-}\rangle & = 24\, \lambda^{(0)}_{2}  D_{1111} \,,\label{0derdfun2}\\
\langle \f_{+} \f_{+} \f_{-} \f_{-}\rangle & = 4\, \lambda^{(0)}_{3}   D_{2211}\,.\label{0derdfun3}
\end{align}
with the corresponding Mellin amplitudes
\begin{align}
    \mathcal{M}_{++++} =&\ \frac{9 \lambda^{(0)}_{1} }{\pi^2}\,,\\
    \mathcal{M}_{----} =&\ \frac{3\lambda^{(0)}_{2} }{\pi^2}\,,\\
    \mathcal{M}_{++--} =&\ \frac{\lambda^{(0)}_{3} }{2\pi^2}\,.
\end{align}
Comparing to the bootstrap results in \eqref{mellinspin01}-\eqref{mellinspin03} results in 
\begin{align}
\lambda^{(0)}_{1}  =&\ \frac{\pi^2}{3}\,a_0\,,\\
\lambda^{(0)}_{2}  =&\ \frac{\pi^2}{3}\,a_0\,,\\
\lambda^{(0)}_{3}  =&-2\pi^2a_0\,.
\end{align}
Plugging this into \eqref{0der} then reproduces the first line of \eqref{shadowpot2}.

\subsection{Spin-1}

Now let us consider spin-1 interactions, which correspond to 2-derivative vertices. The most generic 2-derivative bulk interactions with $\mathbb{Z}_2$ symmetry are given by 
\begin{align}\label{2der}
 V_2(\f_+,\f_-)&=\lambda^{(2)}_{1}  \f_+\f_-\partial_{\mu}\f_-\partial^{\mu}\f^+ + \lambda^{(2)}_{2}  {\f^2_+}\partial_{\mu}\f_+\partial^{\mu}\f_+ +\lambda^{(2)}_{3}  {\f^2_-}\partial_{\mu}\f_-\partial^{\mu}\f_-   \nn & \quad +\lambda^{(2)}_{4} \f^3_+\Box\f_+ +\lambda^{(2)}_{5} \f^3_-\Box\f_- \,.
\end{align}
Using the free equations of motion (EOM)
 \begin{align}
\Box \f_+ =m^2 \f_+\,,\quad  \Box \f_- =m^2 \f_-\,,
\end{align}
and integration by parts (IBP), one can show that there is only one independent vertex, i.e. all the other vertices reduce to that one or a zero-derivative interaction and can therefore be neglected. As an example, let us consider the vertices $\f^2_i  \partial_{\mu}\f_i \partial^{\mu}\f_i$ where $i=\pm$:
\begin{align}
 \int d^4x\, \sqrt{g}\f^2_i \partial_{\mu}\f_i \partial^{\mu}\f_i & =\frac{1}{3} \int d^4x\, \sqrt{g}\, \partial_{\mu}(\f^3_i \partial^{\mu}\f_i)-\frac{1}{3}\int d^4x \,\sqrt{g}\,\f^3_i \Box \f_i \nn
& = -\frac{m^2}{3}\int d^4x \,\sqrt{g} \,\f^4_i\,,
\end{align}
where we have used the EOM for $\f_i$  and IBP in the second line. It follows that the interaction vertices $\f^4_i$ and $\f^2_i \partial_{\mu}\f_i \partial^{\mu}\f_i$ are equivalent. Using similar manipulations on the other vertices, we find that there is only one independent 2-derivative vertex, which is given in \eqref{2derindep}. We can then compute Witten diagrams with four bulk-to-boundary propagators ending on the vertex as before. As shown in Appendix \ref{app2der}, the correlator arising from the potential $V_0+V_2$ is
\begin{align}\label{2dervert}
\langle \f_+\f_+\f_-\f_-\rangle 
&=4\, \lambda^{(0)}_{3}   D_{2211} + \lambda^{(2)}_{1}\bigg(8 D_{2211} + 8(P_1\cdot P_3)D_{3221}+8(P_1\cdot P_4)D_{3212}  \nn
&\qquad +8(P_2\cdot P_3)D_{2321} +8(P_2\cdot P_4)D_{2312}\bigg)\,.
\end{align}
The result consists of pure $D$-functions and scalar products $P_i\cdot P_j$. In Appendix \ref{dmellin}, we will review how to convert such functions to Mellin amplitudes. Following this procedure, the Mellin amplitude of 
\eqref{2dervert} is given by
\begin{align}
    \mathcal{M}_{++--} =&\ \frac{\lambda^{(0)}_{3} }{2\pi^2} -\frac{\lambda^{(2)}_{1}}{8\pi^2}(3s-8)\,.
\end{align}
Comparing this to the bootstrap analysis in Section \ref{spin1_bootstrap}, we conclude that $\lambda^{(2)}_{1}=0$, so spin-1 interactions are forbidden in agreement with the shadow EFT in \eqref{shadowpot2}. 

\subsection{Spin-2}

As explained in previous sections, spin-2 interactions correspond to 4- and 6-derivative interactions. We will consider each in turn.

Modulo EOM and IBP, the most generic 4-derivative interaction vertices are given in \eqref{4derindep}. Using these vertices to compute 4-point Witten diagrams then implies the following correlators \footnote{See \eqref{app:4der} for the generalisation to generic non-identical scalars.}:
\begin{align}
\langle \f_+\f_+\f_+\f_+\rangle & =64\, \lambda^{(4)}_{1} \bigg(3 {D}_{2222} + 4(P_1\cdot P_2) {D}_{3322} +4(P_1\cdot P_3) {D}_{3232} +4 (P_1\cdot P_4) {D}_{3223}  \nn
&\qquad\qquad\quad + 4(P_2\cdot P_3) {D}_{2332} +4 (P_2\cdot P_4) {D}_{2323}+4(P_3\cdot P_4) {D}_{2233} \nn
&\qquad\qquad\quad + 16 (P_1\cdot P_3) (P_2\cdot P_4)\big(1+U + V\big){D}_{3333} \bigg)\,, \label{4dercorr1}\\
\langle \f_-\f_-\f_-\f_-\rangle &=
 8\,\lambda^{(4)}_{2}\bigg(3 {D}_{1111} + 4(P_1\cdot P_2) {D}_{2211} +4(P_1\cdot P_3) {D}_{2121} +4 (P_1\cdot P_4) {D}_{2112}  \nn
&\qquad\qquad\quad + 4(P_2\cdot P_3) {D}_{1221} +4 (P_2\cdot P_4) {D}_{1212}+4(P_3\cdot P_4) {D}_{1122} \nn
&\qquad\qquad\quad + 16 (P_1\cdot P_3) (P_2\cdot P_4)\big(1+U + V\big){D}_{2222} \bigg)\,, \label{4dercorr2}\\
\langle \f_+\f_+ \f_- \f_-\rangle & = 16\, \lambda^{(4)}_{3}\bigg(D_{2211} +4 (P_1 \cdot P_2)   D_{3311} +4  (P_3 \cdot P_4)   D_{2222} + 16 (P_1 \cdot P_2) (P_3 \cdot P_4)  D_{3322}\bigg)\nn
&+ 8 \lambda^{(4)}_{4}\bigg(2 D_{2211} +4 (P_1 \cdot P_3)   D_{3221} +4  (P_2 \cdot P_4)   D_{2312} +4 (P_1 \cdot P_4)   D_{3212} \nn
&\qquad +4  (P_2 \cdot P_3)   D_{2321} + 16 (P_1 \cdot P_3) (P_2 \cdot P_4)(1+V) D_{3322} \bigg)\,. \label{4dercorr3}
\end{align}
Following Appendix \ref{dmellin}, the Mellin transform of \eqref{4dercorr1}-\eqref{4dercorr3} are given by
\begin{align}
    \mathcal{M}_{++++} =&\ \frac{3\lambda^{(4)}_{1}}{4\pi^2} \bigg(\!-8 + \left(35 \left(s^2+t^2+u^2\right)-680\right) \bigg),\\
    \mathcal{M}_{----} =&\  \frac{3\lambda^{(4)}_{2}}{4\pi^2}\left(-\frac{8}{3} + s^2+t^2+u^2\right),\\
    \mathcal{M}_{++--} =&\ \frac{\lambda^{(4)}_{3}}{8\pi^2}\left(64-60 s+15 s^2\right)+\frac{\lambda^{(4)}_{4}}{8\pi^2}\left(151-57s+\frac{15 s^2}{2}-15 t u\right).
\end{align}
Comparing these Mellin amplitudes to the bootstrap results in Section \ref{spin2bootstrap} then fixes the coefficients in \eqref{4derindep} to be (modulo spin-0 solution)
\begin{align}
\lambda^{(4)}_{1} = \frac{4\pi^2}{3}a_1,\quad \lambda^{(4)}_{2}  =  \frac{4\pi^2}{3}a_1,\quad \lambda^{(4)}_{3}  =  -\frac{8\pi^2}{3}a_1,\quad \lambda^{(4)}_{4}  =  -\frac{16\pi^2}{3}a_1,
\end{align}
in perfect agreement with the 4-derivative terms of the shadow action in \eqref{shadowpot2}.

Finally, let us look at 6-derivative interactions. Modulo IBP and EOM, the most general interaction vertices are given in  \eqref{6derindep}. The correlators associated to these vertices are lengthy and can be found in Appendix \ref{6derivativevertex}. After converting to Mellin space and comparing them to the bootstrap results in \eqref{spin2boot}, this constrains the coefficients in \eqref{6derindep} to be 
\begin{align}
\lambda^{(6)}_{1} = \frac{8\pi^2}{45}a_2,\quad \lambda^{(6)}_{2}  =  \frac{8\pi^2}{45}a_2,\quad \lambda^{(6)}_{3}  =  -\frac{16\pi^2}{45}a_2,
\end{align}
which once again reproduces the shadow action in \eqref{shadowpot2} \footnote{Note that the two vertices $\left(\partial\phi_{+}\right)^{2}\left(\nabla^{\mu}\nabla^{\nu}\phi_{-}\right)^{2}$ and $\left(\partial\phi_{-}\right)^{2}\left(\nabla^{\mu}\nabla^{\nu}\phi_{+}\right)^{2}$ in \eqref{shadowpot2} are equivalent under IBP and EOM.}.

\section{Conclusion} \label{conclusion}

The conceptual foundations of holography de Sitter space are far less understood than those of Anti-de Sitter space for a number of reasons. First of all, since the central charge of the boundary theory is tied to cosmological constant, one finds that it should be negative in three boundary dimensions \cite{Anninos:2011ui,Goodhew:2024eup}. In the case of bulk higher spin theories, it is possible to deduce such boundary theories by analytically continuing the AdS/CFT correspondence \cite{Anninos:2011ui}, but it is not clear if such a procedure exists for more realistic bulk theories which reduce to Einstein gravity coupled to lower-spin fields at low energies. Furthermore, one generically expects such a boundary theory to have complex scaling dimensions corresponding to massive states in the bulk which decouple at low energies. Finally, when computing in-in correlators non-perturbatively, we must perform a path integral over all the boundary values of the bulk fields including the metric \cite{Compere:2023ktn,Maldacena:2002vr}. 

In this paper we have asked whether it is possible to derive de Sitter locality from boundary conformal field theory, and found that this is indeed possible. This was demonstrated by mapping in-in correlators to boundary CFT correlators in EAdS$_4$ using the shadow formalism. In contrast to the derivation of AdS locality from CFT found long ago in \cite{Heemskerk:2009pn}, conformal and crossing symmetry are not enough to reconstruct the shadow action because there are two types fields in the bulk after mapping to EAdS. If one starts with a general local EFT of two scalar fields in EAdS, it will not in general correspond to a local EFT in dS unless we impose the additional constraints on the CFT data. We therefore propose additional constraints on anomalous dimensions to fully reconstruct the shadow action. While these constraints can be defined abstractly from a boundary perspective, we check them explicitly from a bulk perspective up to six derivative interactions. To facilitate the analysis, we make use of Mellin space techniques such as dispersion relations which allow for more efficient computation of anomalous dimensions. 

Note that we have only considered conformally coupled scalars in a rigid de Sitter background, so a natural next step would be to consider the more physically interesting case of massless scalars, whose correlators generally exhibit IR divergences. These divergences generically require non-local counterterms \cite{Bzowski:2023nef,Palma:2025oux} although other approaches have been proposed to deal with them such as the stochastic formalism \cite{Starobinsky:1986fx,Gorbenko:2019rza}. On the other hand, inflationary models often include shift symmetries which remove IR divergences \cite{Finelli:2018upr}, so it would be of great interest to investigate whether is possible to reconstruct a massless EFT in dS$_4$ with a shift symmetry from its in-in correlators.  

For the purpose of reconstructing an EFT in dS$_4$, it was sufficient to restrict to contact diagrams in the bulk, but it would also be of interest to understand how to reconstruct tree-level exchange or loop diagrams. For conformally coupled $\phi^4$ theory, renormalised 1-loop correlators were already computed in \cite{Heckelbacher:2022hbq} so this would be a nice starting point. One may hope to achieve this by adapting the AdS bootstrap calculations in \cite{Aharony:2016dwx,Alday:2017gde} and supplementing them with an appropriate generalization of the unmixing and cross-channel constraints found in this paper. Note that coefficients of the shadow EFT can also be fixed by demanding that loop-level in-in correlators are renormalisable (in the case of conformally coupled scalars in the bulk) \cite{Heckelbacher:2022hbq}, or exhibit a certain form of flat space structure \cite{Chowdhury:2023arc,Chowdhury:2025ohm}, so it would be interesting to investigate how these properties can be encoded by the anomalous dimension constraints.

Another natural direction would be to generalise our analysis to gravitational theories in the bulk. Using a generalisation of the shadow formalism to spinning theories \cite{Sleight:2020obc,Sleight:2021plv,Schaub:2023scu,spinninginin}, one can investigate whether a general EFT of gravity  can be reconstructed from a boundary perspective and then use dispersion relations to place bounds on the coefficients following methods developed in AdS \cite{Caron-Huot:2021enk}, which may have observational consequences.

\begin{acknowledgments}

We thank Simon Caron-Huot, Paul Heslop, Ivo Sachs, Charlotte Sleight, and Massimo Taronna for helpful discussions. AL is supported by an STFC Consolidated Grant ST/X000591/1. PD is supported by ANRF Early Career Research Grant ANRF/ECRG/2024/000247/PMS. ZH is supported by National Science Foundation of China under Grant No.~12175197 and Grant No.~12347103.

\end{acknowledgments}

\appendix

\section{Mack polynomials} \label{appx:mack}

We present a formula for Mack polynomials below following \cite{Costa:2012cb} with a slight change of notation \footnote{Our Mack polynomial is $Q_{J,m}(s)$ in equation (176) of \cite{Costa:2012cb}, where $J_\text{there} = \ell$ and $s_\text{there} = -t+\Delta_2+\Delta_3-\tau-2m$. } :
\begin{align}
    \mathcal{Q}_{\tau,\ell,m}(t) =&\ \frac{\mathcal{N}_{\Delta,\ell}}{ \Gamma \left(\frac{\Delta_{1}+\Delta_{2}-(\tau+2m)}{2} \right) \Gamma \left(\frac{\Delta_{3}+\Delta_{4}-(\tau+2m)}{2} \right)}\times Q_{\Delta,\ell,m}(t),  \label{eq:mack1} \\[2mm] 
    Q_{\Delta,\ell,m}(t)=& \sum_{r=0}^{[\ell/2]} (-1)^r\, \frac{\ell! (\ell+\frac{d}{2}-1)_{-r}}{r! (\ell-2r)!} \frac{2^{\ell} (-m)_r\left(-\Delta+\frac{d}{2}-m \right)_r (\ell-2r)!  }{(\Delta-1)_\ell(\Delta+d-1)_\ell} \nn
    & \sum_{\sum k_{ij}=\ell-2r} (-1)^{k_{13}+k_{24}}\prod_{(ij)}\frac{(\delta_{ij})_{k_{ij}}}{k_{ij}!} \prod_{n=1}^4 \left( \alpha_n   \right)_{\ell-r-\sum_j k_{jn}}, \label{eq:mack2} \\[2mm]
    \mathcal{N}_{\Delta,\ell} =&\ \frac{2^{-\ell} \Gamma (\Delta+\ell ) (\Delta -1)_\ell}{m! (\Delta-\frac{d}{2} +1)_m \Gamma \left(\frac{\Delta+\ell -\Delta_{12}}{2}\right) \Gamma \left(\frac{\Delta+\ell +\Delta_{12}}{2} \right) \Gamma \left(\frac{\Delta+\ell -\Delta_{34}}{2} \right) \Gamma \left(\frac{\Delta+\ell +\Delta_{34}}{2} \right)}. \label{eq:mack3}
\end{align}
In this expression the label $(ij)$ takes values in the set $\{(13),(14),(23),(24)\}$. The variables $\delta_{ij}$ and $\alpha_n$ are 
\begin{align}
    \delta_{13}=&\ \frac{ t+\tau+2m-\Delta_2-\Delta_4}{2}\,,& \delta_{14}=&\ \frac{\Delta_1+\Delta_4-t}{2}\,,\\
    \delta_{24}=&\ \frac{ t+\tau+2m-\Delta_1-\Delta_3}{2}\,,& \delta_{23}=&\ \frac{\Delta_2+\Delta_3-t}{2}\,, \\
    \alpha_{1}=&\ 1+\frac{\Delta-\ell-d-\Delta_{12}}{2}\,,& \alpha_{2}=&\  1+\frac{\Delta-\ell-d+\Delta_{12}}{2}\,, \\
    \alpha_{3}=&\ 1-\frac{\Delta +\ell+\Delta_{34}}{2}\,,& \alpha_{4}=&\ 1-\frac{\Delta +\ell-\Delta_{34}}{2}\,,
\end{align}
where $\Delta_{ij}=\Delta_i-\Delta_j$.

\section{Unmixing matrix} \label{unmixing}
The procedure for unmixing degenerate CFT operators was fist developed in the context of correlation functions in $\mathcal{N}=4$ SYM \cite{Aprile:2017xsp}. For the sake of being self-contained we will review that procedure in this Appendix. Although in the present context there are only two single-trace operators $\mathcal{O}_+$ and $\mathcal{O}_-$, we will keep the discussion general and assume there exist a family of single-trace operators labeled by $\mathcal{O}_p$ (where $p=\pm$ in our context). Following a similar discussion to Section \ref{sec:adcon}, the double-trace operators 
\begin{align}
    [\mathcal{O}_p\mathcal{O}_q]_{n,\ell} \equiv\  :\mathcal{O}_p \square^n \partial^\ell \mathcal{O}_q:
\end{align}
might mix with other double-trace operators with the same spin and classical twist. The eigenstates of the dilatation operator $[\mathcal{O}\mathcal{O}]^i_{n,\ell}$ are linear combinations of $[\mathcal{O}_p\mathcal{O}_q]_{n,\ell}$
\begin{align}\label{eq:basis}
    [\mathcal{O}\mathcal{O}]^i_{n,\ell} = \sum_{p,q} R_{i,(pq)} [\mathcal{O}_p\mathcal{O}_q]_{n,\ell}\,.
\end{align}
Note that \eqref{eq:basis} is simply a change of basis, and the total number of $[\mathcal{O}\mathcal{O}]^i_{n,\ell}$ and  $[\mathcal{O}_p\mathcal{O}_q]_{n,\ell}$ are the same for a given twist and spin. We can thus view $i$ and $\alpha=(pq)$ as the vector indices of two bases. If we normalize these operators in the usual way (omitting spacetime dependence)
\begin{align}
    \langle\, [\mathcal{O}\mathcal{O}]^i [\mathcal{O}\mathcal{O}]^j \rangle = \delta_{ij},\qquad \langle [\mathcal{O}_p\mathcal{O}_q] [\mathcal{O}_r\mathcal{O}_s] \rangle = \delta_{(pq),(rs)} \equiv \delta_{\alpha\beta}\,,
\end{align}
the matrix $R_{i\alpha}$ will be an orthogonal matrix, since the two-point functions can be thought of as inner products of basis vectors and $R_{i\alpha}$ connects two orthonormal bases. 

Now we consider the averaged CFT data obtained by summing over all dilatation eigenstates:
\begin{align}
    \langle a^{(0)} \rangle_{pqrs} = \sum_i C^{(0)}_{pq,i} C^{(0)}_{rs,i}\,,\qquad  \langle a^{(0)}\gamma^{(1)} \rangle_{pqrs} = \sum_i C^{(0)}_{pq,i}\, \gamma^{(1)}_i C^{(0)}_{rs,i}\,.
\end{align}
These objects can be viewed as matrices whose elements are labeled by $\alpha=(pq)$ and $\beta=(rs)$:
\begin{align}
    \langle a^{(0)} \rangle_{\alpha\beta} = \sum_i C^{(0)}_{\alpha i} C^{(0)}_{\beta i}\,, \qquad  \langle a^{(0)}\gamma^{(1)} \rangle_{\alpha\beta} = \sum_i C^{(0)}_{\alpha i}\, \gamma^{(1)}_i C^{(0)}_{\beta i}\,,
\end{align}
which can be written in terms of matrix products as follows: 
\begin{align}
    \langle a^{(0)} \rangle = C \cdot C^T\,, \qquad \langle a^{(0)}\gamma^{(1)} \rangle = C \cdot \Gamma \cdot C^T\,,
    \label{cftmatrices1} 
\end{align}
where
\begin{align}  
    C\equiv C_{\alpha i} = C^{(0)}_{\alpha i}\,, \qquad \Gamma \equiv \Gamma_{ij} = \gamma^{(1)}_i\delta_{ij}\,.
\end{align}
Our goal is to solve for the anomalous dimensions $\gamma^{(1)}_i$ in $\Gamma$ in terms of the matrices $\langle a^{(0)} \rangle$ and $\langle a^{(0)}\gamma^{(1)} \rangle$. 

At the disconnected level we can use both $[\mathcal{O}\mathcal{O}]^i_{n,\ell}$ and $[\mathcal{O}_p\mathcal{O}_q]_{n,\ell}$ as the basis to represent $\langle a^{(0)} \rangle$. When changing to the basis of $[\mathcal{O}_p\mathcal{O}_q]_{n,\ell}$, we have 
\begin{align}
    \langle a^{(0)} \rangle = \widetilde{C} \cdot \widetilde{C}^T,\qquad \widetilde{C} = C \cdot R\, .
\end{align}
However, $\widetilde{C}$ encodes three-point functions
\begin{align}
    \langle \mathcal{O}_{p}\mathcal{O}_q [\mathcal{O}_r\mathcal{O}_s]_{n,\ell} \rangle \propto \widetilde{C}_{pq,(rs)}\,,
\end{align}
and is non-zero only when $(pq)=(rs)$. Therefore $\widetilde{C}$ is in fact a diagonal matrix, as well as $\langle a^{(0)} \rangle$. Using this property, we can solve $\widetilde{C}$ in terms the square root of $\langle a^{(0)} \rangle$ to obtain 
\begin{align}
    \widetilde{C} = \langle a^{(0)} \rangle^{\frac{1}{2}},\qquad C = \widetilde{C}\cdot R^{-1} = \langle a^{(0)} \rangle^{\frac{1}{2}}\cdot R^{-1}.
\end{align}
Substituting this into \eqref{cftmatrices1} then implies that 
\begin{align}
    \Gamma = R\cdot \langle a^{(0)} \rangle^{-\frac{1}{2}} \cdot \langle a^{(0)}\gamma^{(1)} \rangle \cdot \langle a^{(0)} \rangle^{-\frac{1}{2}} \cdot R^{-1}.
\end{align}
Although the explicit form of $R$ is unknown, it is an orthogonal matrix, therefore the two matrices $\Gamma$ and $\langle a^{(0)} \rangle^{-\frac{1}{2}} \cdot \langle a^{(0)}\gamma^{(1)} \rangle \cdot \langle a^{(0)} \rangle^{-\frac{1}{2}}$ share the same set of eigenvalues. The eigenvalues of $\Gamma$ are exactly the anomalous dimensions $\gamma^{(1)}_i$, which means that  
\begin{align}
    \{ \gamma^{(1)}_i \} = \{\text{eigenvalues of $M\equiv \langle a^{(0)} \rangle^{-\frac{1}{2}} \cdot \langle a^{(0)}\gamma^{(1)} \rangle \cdot \langle a^{(0)} \rangle^{-\frac{1}{2}}$}\}.
\end{align}
Writing out each element of $M$ explicitly, we obtain \eqref{eq:mat}. 

\section{Spin-2 CFT data}\label{spin2}

In this Appendix we provide more details for the bootstrap analysis in Section \ref{spin2bootstrap}. In particular, the unmixing matrix elements are 
\begin{align}
    M^{--}_{n,\ell} =&\ \left(a_0 + \frac{4}{9}\left(20 n^4+100 n^3+199 n^2+185 n+78\right)  a_1 \right. \nn
    & \left. \quad + \frac{4}{45} \left(2 n^2+5 n+6\right)\left(4 n^4+20 n^3+27 n^2+5 n+4\right) a_2 \right)\delta_{\ell,0} \nn 
    &+ \left(\frac{4}{45}(n+2) (n+3) (2 n+3) (2 n+5) a_1 \right. \nn 
    & \left. \quad -\frac{4}{225} (n+2) (n+3) (2 n+3) (2 n+5) \left(2 n^2+9 n+12\right) a_2 \right) \delta_{\ell,2},  
\end{align}
\begin{align}    
    M^{++}_{n,\ell} =&\ \left(\frac{b_0}{3}  + \frac{4}{315} \left(20 n^4+100 n^3+199 n^2+185 n+588\right)  b_1 \right. \nn
    & \left. \quad + \frac{4}{2835}\left(8 n^6+60 n^5+58 n^4-335 n^3-999 n^2-1060 n+4536\right) b_2 \right)\delta_{\ell,0} \nn 
    &+ \left(\frac{4}{1575}(n+2) (n+3) (2 n+3) (2 n+5) b_1 \right. \nn 
    & \left. \quad -\frac{4}{14175}(n+2) (n+3) (2 n+3) (2 n+5) \left(2 n^2+9 n+18\right) b_2 \right) \delta_{\ell,2},
\end{align}
\begin{align}    
    M^{+-}_{n,\ell} =&\ \left(c_0  + \frac{2}{3} \left(2 n^2+5 n+6\right) c_1 + \frac{4}{15} \left(4 n^4+20 n^3+61 n^2+90 n+60\right)c_2 \right. \nn
    & \left. \quad + \frac{1}{45} \left(8 n^4+40 n^3-14 n^2-160 n+81\right) c_3 \right. \nn 
    &\left. \quad + \frac{2}{315} \left(16 n^6+120 n^5+252 n^4+10 n^3-562 n^2-655 n+1134\right)c_4 \right)\delta_{\ell,0} \nn 
    &+ \left(-\frac{2}{225} (n+2) (n+3) (2 n+3) (2 n+5) c_3  \right. \nn 
    & \left. \quad -\frac{4}{1575}  (n+2) (n+3) (2 n+3) (2 n+5) \left(2 n^2+9 n+14\right) c_4 \right) \delta_{\ell,2}.
\end{align}
Moreover the t-channel anomalous dimension is 
\begin{align}
    \gamma_\text{pure} =&\ \left(-2c_0  + \frac{2}{3} \left(2 n^2+3 n-4\right) c_1 -\frac{4}{45} \left(8 n^4+24 n^3-32 n^2-75 n+30\right)c_2 \right. \nn
    & \left. \quad + \frac{2}{15} \left(8 n^4+24 n^3+16 n^2-3 n-75\right) c_3 \right. \nn 
    &\left. \quad -\frac{2}{315} \left(32 n^6+144 n^5+56 n^4-372 n^3-790 n^2-645 n+2520\right) c_4 \right)\delta_{\ell,0} \nn 
    &+ \left(\frac{2}{9} (n+1) (2 n+3) c_1 -\frac{4}{45} (n-1) (n+1) (2 n+3) (2 n+7) c_2 \right. \nn 
    & \left. \quad -\frac{2}{45} (n+1) (2 n+3) \left(4 n^2+10 n+15\right) c_3 + \frac{2}{315} (n+1) (2 n+3) \left(4 n^2+10 n+21\right)c_4 \right) \delta_{\ell,1}, \nn
    &+ \left(-\frac{4}{225} (n+1) (n+2) (2 n+3) (2 n+5) c_2  \right. \nn 
    & \left. \quad + \frac{4}{1575}(n+1) (n+2) (2 n+3) (2 n+5) \left(4 n^2+14 n+21\right) c_4 \right) \delta_{\ell,2}.
\end{align}

\section{Details of bulk calculations}\label{positionsp}
In this section we present formulae for contact Witten diagrams describing higher-derivative interactions of four non-identical scalars $\phi_i\ (i=1,2,3,4)$ whose dual operators have scaling dimensions $\D_i \ (i=1,2,3,4)$. Note that if some of the scalars are identical we need to include extra channels.

\subsection*{Two-derivative vertex}\label{app2der}
For the two-derivative vertex $\partial_\mu\f_1 \partial^\mu\f_2 \ \f_3 \f_4 $ we obtain
\begin{align}
& \int_{AdS} dX\ \nabla_{\mu}G^{\D_1}_{B\partial}(P_1, X)\ \nabla^{\mu}G^{\D_2}_{B\partial}(P_2, X)\ G^{\D_3}_{B\partial}(P_3, X) G^{\D_4}_{B\partial}(P_4, X)\nn
&=\D_1 \D_2 D_{\D_1\D_2\D_3\D_4}+4 \D_1\D_2\, (P_1\cdot P_2)\, D_{\D_1+1\D_2+1\D_3\D_4}\,.
\end{align}

\subsection*{Four-derivative vertex}\label{app:4der}
For the four-derivative vertex $ \partial_\mu\f_1 \partial^\mu\f_2 \partial_\nu\f_3 \partial^\nu\f_4$ we obtain
\begin{align}\label{db4der}
&\int_{AdS} dX \ \nabla_{\mu}G^{\D_1}_{B\partial}(P_1,X) \ \nabla^{\mu}G^{\D_2}_{B\partial}(P_2,X)\ \nabla_{\nu}G^{\D_3}_{B\partial}(P_3,X) \  \nabla^{\nu}G^{\D_4}_{B\partial}(P_4,X)\nn 
&=  \D_1\D_2\D_3\D_4 \bigg( D_{\D_1 \D_2 \D_3 \D_4} +4 (P_1 \cdot P_2) \,  D_{\D_1+1\, \D_2+1\, \D_3 \D_4} +4  (P_3 \cdot P_4)  \, D_{\D_1 \D_2 \D_3+1\, \D_4+1}\nn
&\quad  + 16 (P_1 \cdot P_2) (P_3 \cdot P_4) \, D_{\D_1+1\, \D_2+1\, \D_3+1\, \D_4+1} \bigg)\,.
\end{align}

\subsection*{Six-derivative vertex} \label{6derivativevertex}
For the six-derivative vertex $ \partial_\mu\f_1 \partial^\mu\f_2\  \nabla_\nu \nabla_\rho \f_3 \nabla^\nu \nabla^\rho\f_4$ we obtain
\begin{align}\label{app6der}
&\int_{AdS} dX \ \nabla_{\mu}G^{\D_1}_{B\partial}(P_1,X)\nabla^{\mu}G^{\D_2}_{B\partial}(P_2,X)\nabla_{\nu}\nabla_{\rho}G^{\D_3}_{B\partial}(P_3,X) \ \nabla^{\nu}\nabla^{\rho}G^{\D_4}_{B\partial}(P_4,X)\nn  
&= \D_1 \D_2\D_3\D_4 \bigg((\D_3\D_4+2)D_{\Delta _1\Delta _2\Delta _3\Delta _4}+4 (\D_3\D_4+2) (P_1 \cdot P_2) D_{\Delta _1+1\,\Delta _2+1\,\Delta _3\Delta _4}\nn
&\qquad +4 (P_3 \cdot P_4) (1+2\D_3+2\D_4+2\D_3\D_4)D_{\Delta _1\Delta _2\Delta _3+1\,\Delta _4+1}\nn
&\qquad+ 16 (P_1\cdot P_2) (P_3 \cdot P_4) (1+2\D_3+2\D_4+2\D_3\D_4)D_{\Delta _1+1\,\Delta _2+1\,\Delta _3+1\,\Delta _4+1}\nn
&\qquad+16 (1+\D_3)(1+\D_4) (P_3 \cdot P_4)^2 D_{\Delta _1\Delta _2\Delta _3+2\,\Delta _4+2}\nn
&\qquad+ 64(1+\D_3)(1+\D_4) (P_1 \cdot P_2) (P_3 \cdot P_4)^2 D_{\Delta _1+1\,\Delta _2+1\,\Delta _3+2\,\Delta _4+2}\bigg)\,.
\end{align}

\section{Converting correlators into Mellin space}\label{dmellin}

For correlators like \eqref{2dervert} which have both $D$-functions and scalar products $P_i\cdot P_j$, the Mellin amplitudes generally depend on the Mellin variables $s$ and $t$. When written as functions of cross-ratios, such correlators are linear combinations of $U^m V^n D_{\{\Delta'_i\}}$:
\begin{align}
    \mathcal{G}_{\{\Delta_i\}}(z,\zb) = \sum a_{m,n,\{\Delta'_i\}} U^m V^n D_{\{\Delta'_i\}}(z,\zb)\,.
\end{align}
To convert this to Mellin space, we consider the Mellin transformation of each term. For $U^m V^n D_{\{\Delta'_i\}}$, following \eqref{eq:Dmellin}, we can write 
\begin{align}
    U^m V^n D_{\{\Delta'_i\}} = \frac{\pi ^{3/2}  \Gamma \left(\frac{\Delta'_1+\Delta'_2+\Delta'_3+\Delta'_4}{2}\!-\!\frac{3}{2}\right)}{2 \Gamma (\Delta'_1) \Gamma (\Delta'_2) \Gamma (\Delta'_3) \Gamma (\Delta'_4)}\! \int \frac{\dd s' \dd t'}{(2\pi i)^2} U^{\frac{s'+2m}{2}} V^{\frac{t'+2n-\Delta'_2-\Delta'_3}{2}}  \Gamma_{\{\Delta'_i\}}(s',t').
\end{align}
However, the transformation kernel is not the same as one in \eqref{eq:mellindef}. To obtain the correct kernel, we need to take
\begin{align}
    s'=s-2m\,, \qquad t' = t-2n+\Delta'_2+\Delta'_3-\Delta_2-\Delta_3\,.
\end{align}
The Mellin amplitude of $U^m V^n D_{\{\Delta'_i\}}$ is then given by 
\begin{align}
    \mathcal{M}_{m,n,\{\Delta'_i\}}(s,t) =&\ \frac{\pi ^{3/2}  \Gamma \left(\frac{\Delta'_1+\Delta'_2+\Delta'_3+\Delta'_4}{2}-\frac{3}{2}\right)}{2 \Gamma (\Delta'_1) \Gamma (\Delta'_2) \Gamma (\Delta'_3) \Gamma (\Delta'_4)} \frac{\Gamma_{\{\Delta'_i\}}(s',t')}{\Gamma_{\{\Delta_i\}}(s,t)} \nn
    =&\ \frac{\pi ^{3/2}  \Gamma \left(\frac{\Delta'_1+\Delta'_2+\Delta'_3+\Delta'_4}{2}-\frac{3}{2}\right)}{2 \Gamma (\Delta'_1) \Gamma (\Delta'_2) \Gamma (\Delta'_3) \Gamma (\Delta'_4)} \frac{\Gamma_{\{\Delta'_i\}}(s-2m,t-2n+\Delta'_2+\Delta'_3-\Delta_2-\Delta_3)}{\Gamma_{\{\Delta_i\}}(s,t)},
\end{align}
and the whole result is 
\begin{align}
    \mathcal{M}_{\{\Delta_i\}}(s,t) = \sum a_{m,n,\{\Delta'_i\}} \mathcal{M}_{m,n,\{\Delta'_i\}}(s,t)\,.
\end{align}

\bibliographystyle{JHEP}
\bibliography{ds}

\end{document}